\documentclass[11pt]{article}

\usepackage[a4paper, margin=0.8in]{geometry}

\usepackage{graphicx} % Required for inserting images
\usepackage{booktabs}
\usepackage{amsmath}
\usepackage{url} 
\usepackage{natbib}

\usepackage{amsmath, amssymb}
\usepackage{bm}
\usepackage{xcolor}
\usepackage{tabularx}

\usepackage{makecell} % in preamble
\usepackage{pifont}

\newcommand{\cmark}{\ding{51}}       % checkmark
\newcommand{\xmark}{\ding{55}}       % x mark
\newcommand{\opencircle}{\ding{109}} % open circle

\title{Areal Disaggregation: A Small Area Estimation Perspective\footnote{This manuscript has been authored by UT-Battelle, LLC, under contract DE-AC05-00OR22725 with the US Department of Energy (DOE). The US government retains and the publisher, by accepting the article for publication, acknowledges that the US government retains a nonexclusive, paid-up, irrevocable, worldwide license to publish or reproduce the published form of this manuscript, or allow others to do so, for US government purposes. DOE will provide public access to these results of federally sponsored research in accordance with the DOE Public Access Plan (https://www.energy.gov/doe-public-access-plan).}}

\author{by Yunhan Wu$^1$, Finn Lindgren$^2$ and Heidi A. Hanson$^{1}$\\
\vspace{0.2em}\\
$^1$Oak Ridge National Laboratory, Oak Ridge, TN, 37830, USA\\ 
$^2$School of Mathematics and Maxwell Institute of Mathematical Sciences,\\ University of Edinburgh, Edinburgh, UK\phantom{te}}

\date{}

\begin{document}

\maketitle

\abstract{Producing reliable estimates of health and demographic indicators at fine areal scales is crucial for examining heterogeneity and supporting localized health policy. However, many surveys release outcomes only at coarser administrative levels, thereby limiting their relevance for decision-making. We propose a fully Bayesian, single-stage spatial modeling framework for area-level disaggregation that generates fine-scale estimates of indicators directly from coarsely aggregated survey data. By defining a latent spatial process at the target resolution and linking it to observed outcomes through an aggregation step, the framework adopts small-area estimation techniques while incorporating covariates and delivering coherent uncertainty quantification. The proposed methods are implemented with \texttt{inlabru} to achieve computational efficiency. We evaluate performance through a simulation study of general fertility rates in Kenya to demonstrate the models’ ability to recover fine-scale variation across diverse data-generating scenarios. We further apply the framework to two national surveys to produce district-level fertility estimates from the 2022 Kenya Demographic and Health Survey and, more importantly, district-level indicators for unpaid care and domestic work and mass media usage from the 2021 Kenya Time Use Survey.}

\section{Introduction}

Small area estimation (SAE) plays an important role in modern public health and epidemiology in the production of reliable estimates for geographic subpopulations in which direct survey data are sparse or unavailable. With a growing emphasis on localized planning and addressing health inequalities, there is a clear demand for estimates of disease burden and health indicators at fine-areal scales. For example, in low- and middle-income countries (LMICs), this often corresponds to the district (Admin-2) level, which typically serves as the operational unit for management decisions and resource allocation. Empirical studies and international initiatives---in particular the Sustainable Development Goals (SDG) effort---have explicitly called for reporting health indicators at an area-level granularity \citep{oosterhof2018localizing,janocha2021guidance,jonsson2021localizing}. Studies have corroborated these needs across specific indicators, including (but not limited to) vaccination coverage \citep{utazi:etal:21}, educational attainment \citep{delprato2024spatial,wu2025education}, fertility \citep{abate2024mapping}, and child mortality \citep{wu:etal:21}.

Meeting these needs requires statistical approaches capable of generating robust estimates when direct survey data are insufficient. Design-based SAE methods, such as the Horvitz-Thompson estimator \citep{horvitz1952generalization}, are limited when handling small sample sizes for many areas of interest. To address this, model-based SAE methods have become increasingly popular \citep{rao2015small}. Classic area-level models, such as the Fay--Herriot (FH) model \citep{fay1979estimates}, and unit-level models, such as the one developed by \citet{battese1988error}, allow for borrowing statistical strength across areas by incorporating auxiliary information, typically from censuses or administrative records \citep{pratesi2016analysis, morales2021model, moretti2021comparison}. The Bayesian formulation of these methods comes naturally with uncertainty quantification and facilitates hierarchical modeling that incorporates covariates and spatial or temporal dependence structures. %\textcolor{red}{add continuous SPDE}

Despite these advances, SAE frameworks often assume that response data and the target level of inference are defined at the same geographic resolution. For example, many demographic and health surveys (DHS) provide GPS coordinates that allow individuals to be assigned to districts (Admin-2). In many other cases, however, survey outcomes are released only at coarser spatial scales owing to confidentiality considerations or data collection constraints, whereas policy and research applications require estimates at finer scales.

Consider the 2021 Kenya Time Use Survey (KTUS) \citep{KNBS2021TimeUse}. The publicly released data are only geo-indexed at the county (Admin-1) level, even though many key indicators would ideally be mapped at the district scale to guide localized interventions. This limitation is especially relevant for SDG 5.4.1 \citep{UNSD_SDG541}, which tracks the proportion of time spent on unpaid domestic and care work by sex, age, and location. District-level estimates are essential for uncovering heterogeneity and informing targeted household welfare and labor policies. Without methodological innovations, however, such coarse data cannot be leveraged for fine-scale inference.

% This misalignment motivates spatial downscaling (or disaggregation) methods: approaches that infer high-resolution risk maps or prevalence surfaces from aggregated data, by leveraging auxiliary geospatial covariates and spatial models.

%\textcolor{red}{Mention geospatial models here which automatically do downscaling with points. Make clear we are dealing with areal data to start with}

%\textcolor{red}{Motivating example on Kenya Time report survey}

%\subsection*{From Small Area Estimation to Spatial Downscaling}

Although SAE methods are designed to address the challenge of small sample sizes within predefined geographic units, disaggregation methods tackle a different but related problem: generating predictions at spatial resolutions that are finer than those of the observed (aggregated) data. This challenge arises across fields ranging from geography, remote sensing, demography, and epidemiology and is referred to under various names, including downscaling, disaggregation, dasymetric mapping, or super-resolution. Despite differences in framing, these approaches aim to enhance spatial detail by using auxiliary data and modeling assumptions.

A widely used approach is dasymetric mapping, which redistributes areal data into smaller zones by using ancillary covariates such as land cover, building density, or remote sensing proxies. Population mapping products such as LandScan \citep{dobson2000landscan} and WorldPop \citep{tatem2017worldpop}, as well as environmental health applications \citep{requia2018modeling}, are prominent examples. However, these methods typically treat aggregate totals as fixed and ignore sampling or measurement error. Although effective for stable quantities such as population counts, they are less suited to outcomes derived from surveys, in which uncertainty must be explicitly propagated. In such cases, SAE models are better equipped for the task.

Another strategy is spatial microsimulation, which constructs synthetic microdata constrained to match observed marginals. The approach parallels multilevel regression with poststratification (MRP)\citep{park2004bayesian}: outcomes are modeled by demographic group and then redistributed via synthetic population composition. Applications span health, welfare, transport, and demography \citep{tanton2014review,smith2021estimating}. However, microsimulation faces challenges of computational scaling, validation, and uncertainty quantification.

A different class of approaches employs geostatistical models to estimate continuously indexed spatial surfaces, often via the stochastic partial differential equation framework \citep{lindgren2011explicit}. These models link aggregated data to a continuous latent spatial field and are widely used in disease mapping, including COVID-19 studies \citep{python2022high}. They provide a coherent Bayesian framework with uncertainty quantification and are supported by software such as the disaggregation package in R \citep{nandi2023disaggregation}. However, their performance is highly dependent on high-quality spatial covariates, which are not always predictive of health outcomes. Moreover, validation at fine resolutions remains challenging because of the lack of ground-truth data \citep{arambepola2020nonparametric}.

Importantly, many of these disaggregation methods do not account for survey design, which is a core feature of survey data for health and demographic indicators. SAE methods are uniquely positioned to bridge this gap by combining design-based consistency with model-based flexibility.

Given the limitations of the existing methods and the practical goal of generating reliable area-level estimates, we formulate a novel area-level SAE approach for spatial disaggregation of health indicators. Our approach is formulated as a fully Bayesian, single-stage spatial model that defines a latent process at the target resolution and incorporates information from area-level covariates. An inherent aggregation step links this latent field to the outcomes observed at coarser scales to ensure consistency between the observed survey data and the finer level of inference. Although our motivating examples come from LMIC contexts, the proposed framework is broadly applicable across diverse contexts. This formulation allows the framework to accommodate classical SAE methods, such as the FH and unit-level models, and can also incorporate MRP when demographic microdata are available. Computationally, the models are implemented by using \texttt{inlabru} and related tools to achieve computational efficiency \citep{bachl2019inlabru,lindgren2024inlabru,suen2025influence}. By construction, the framework delivers coherent uncertainty quantification and enables standard Bayesian inference.

The remainder of this paper is organized as follows. Section~\ref{sec:method} describes the methodological framework: it begins with a review of existing SAE methods and then introduces our proposed area-level disaggregation models. Section~\ref{sec:simulation} presents a simulation study on general fertility rate~(GFR) in Kenya to evaluate model performance. Sections~\ref{sec:case-study-1} and \ref{sec:case-study-2} apply the framework to two real-world case studies---the 2022 Kenya DHS and the KTUS. Finally, Section~\ref{sec:discussion} discusses the findings and their implications.

\section{Method}
\label{sec:method}
\subsection{Overview of SAE Models}

This section describes three core classes of methods for SAE: direct estimation, area-level (FH) models, and unit-level models \citep{rao2015small,fay1979estimates,battese1988error}. These approaches assume that survey data are geo-indexed at the level of interest, which means that outcomes are observed for sampled individuals in the finer subareas rather than only in aggregated form. Later in this work, we extend beyond these standard frameworks to introduce our areal disaggregation approach to address the more challenging case of when data are available only at coarser geographic levels.

These methods form a natural spectrum: from design-based estimators that rely solely on the sampling design, to increasingly model-based approaches that trade some design consistency for greater flexibility in handling sparse data. In addition to these three approaches, we also consider MRP \citep{downes2018mrp}, which can be combined with area-level or unit-level models to incorporate additional demographic features and align estimates with the population structure. MRP is an extension that partitions the population into cells defined by demographic variables, models the outcomes within each cell, and then aggregates predictions with known population counts.

% In this section we review three standard classes of methods for SAE (citations needed). These approaches assume that the survey data are geo-indexed at the level of interest, meaning that outcomes are observed for sampled individuals in the finer subareas rather than only in aggregated form. Under this setting, the methods move from purely design-based survey weighted estimators, which rely only on the sampling design, to model-based estimators, which trade some design robustness for increased flexibility in handling sparse data. In a later section, we extend beyond this setting and introduce our proposed areal disaggregation framework, which addresses the more challenging case where data are available only at coarser geographic levels.

Consider a finite target population with $N$ individuals living in $|\mathcal{I}|$ coarse areas (e.g., Admin-1) indexed by $i = 1, \dots, |\mathcal{I}|$. Each area $i$ contains a set of finer subareas (e.g., Admin-2), which we index by $j = 1,\dots,|\mathcal{J}|$. We use $i[j]$ to denote the parent coarser area that contains subarea $j$. Within each subarea $j$, there are $N_j$ individuals indexed by $k = 1, \dots, N_j$.

Let $y_{jk}$ denote the outcome for individual $k$ in subarea $j$. Our methodological framework focuses on two outcome types of primary interest in SAE for health indicators: binary and Poisson outcomes. The binary case ($y_{jk} \in \{0,1\}$) is common in disease mapping and corresponds to prevalence estimation. The Poisson case ($y_{jk} \in \{0,1,2,\dots\}$) is appropriate for rare events or rate-type quantities such as fertility or mortality rates. 

Our primary objective is to estimate the prevalence or rate at the subarea level, defined as
\[
\mu_j = \frac{1}{N_j} \sum_{k=1}^{N_j} y_{jk},
\]
where $N_j$ is the population size of subarea $j$. 

Although not of direct interest, we also define the prevalence or rate at the coarse-area level as a population-weighted aggregation of its subareas. This formulation motivates the disaggregation approaches introduced in later sections:
\begin{equation*}
\mu_i =  \sum_{j : i[j] = i} \frac{N_j}{N_i} \mu_j ,
%\label{eq:area-subarea-population-aggre}
\end{equation*}
where $N_i =\sum_{j : i[j] = i} N_{j}$ is the population in area $i$.

An important aspect of SAE is the survey sampling design. Individuals are typically sampled through complex designs that involve unequal probabilities of selection. For each subarea $j$, a sample of size $n_j$ is drawn according to the survey design, which we denote as $S_j \subset \{y_{jk} : k=1,\dots,N_j\}$ with $|S_j| = n_j$. For each sampled individual $k \in S_j$, the associated design weight $w_{jk}$ is defined as the inverse of its inclusion probability such that $w_{jk} = \frac{1}{\pi_{jk}}$ with $ \pi_{jk} = \Pr(k \in S_j)$.

Given this setup, we describe three standard approaches for SAE when data are geographically referenced at the subarea level.

\subsubsection{Direct Estimation}
\label{sec:direct-model-admin2}
Direct estimators use only the sampled data within each subarea and account for the survey design through weights \citep{rao2015small}. A weighted estimator of prevalence in subarea $j$ is given by \citep{hajek:71}:
\[
\hat \mu_j^w = \frac{\sum_{k \in S_j} w_{jk} y_{jk}}{\sum_{k \in S_j} w_{jk}}.
\]
These estimates are design-consistent and require minimal modeling assumptions, but they often have large variances when sample sizes are small. This motivates the use of model-based approaches, which borrow strength from neighboring regions.

\subsubsection{Area-Level Models}
\label{sec:FH-model-admin2}
Area-level models, or FH models \citep{fay1979estimates}, extend direct estimation by linking areas through a hierarchical model. Let $\hat \mu_j^w$ denote the direct estimate of prevalence or rate in subarea $j$ and define the transformed estimate as
\[
\hat \lambda_j^w = g(\hat \mu_j^w), \qquad V_j = \widehat{\mathrm{var}}(\hat \lambda_j^w),
\]
where $g(\cdot)$ is the link function or logit and log transformation for prevalence and rates, respectively. The variance $V_j$ on the transformed scale can be obtained from the design-based variance by using the Delta method.

The FH model assumes
\begin{align}
\begin{split}
\hat \lambda_j^w \mid \lambda_j &\sim \mathcal{N}(\lambda_j, V_j), \\
\lambda_j &= \alpha + \bm{x}_j^\top \bm{\beta} + b_j,
\label{eq:FH-adm2-eta}
\end{split}
\end{align}

where $\bm{x}_j$ are subarea-level covariates, $\bm{\beta}$ are their coefficients, and $b_j$ are random effects. The prevalence or rate in subarea $j$ is then $\mu_j = g^{-1}(\lambda_j)$.

To capture spatial dependence in $b_j$, we adopt the BYM2 formulation \citep{riebler2016an}, which is a reparameterization of Besag-York-Mollie (BYM) \citep{besag1991bayesian} and decomposes the random effect into an unstructured independent and identically distributed (IID) term and a spatially structured intrinsic conditional autoregressive component. The total variance and the proportion attributable to the structured component are governed by hyperparameters, for which we specify penalized complexity priors \citep{simpson2017penalising}. Full details, including the hyperprior setup, are provided in Section \ref{sec:BYM2-full-specification} of the supplemental material. We apply the same spatial structure across all models discussed later.

Smoothing reduces variance by borrowing strength across subareas, but FH models still depend on stable direct estimates and their variances, which can limit performance when data are very sparse.

% Posterior inference is conducted using the Integrated Nested Laplace Approximation (INLA) method, implemented in the R package \texttt{INLA} \citep{rue2009approximate}. 

% FH models reduce variance by borrowing strength across subareas, but rely on stable direct estimates and their variances, which can limit performance when data are very sparse.

\subsubsection{Unit-Level Models}
\label{sec:unit-model-admin2}

Unit-level SAE models directly use individual outcomes and are typically implemented at the cluster level in surveys with cluster sampling. For a survey with $C$ clusters, let $Y_c$ and $n_c$ denote the number of positive outcomes and the number of sampled individuals in cluster $c$, respectively. Depending on the outcome type, the data are modeled as
\[
Y_c \mid \mu_c \sim 
\begin{cases}
\text{Binomial}(n_c, \mu_c), & \text{binary outcomes}, \\
\text{Poisson}(n_c \mu_c), & \text{count outcomes},
\end{cases}
\]
where $\mu_c$ represents the underlying prevalence or rate in cluster $c$.

% The mean structure is specified through a link function:
% \begin{equation}
% g(\mu_c) = \alpha + \bm{x}_{j[c]}^\top \bm{\beta} + b_{j[c]},
% \label{eq:unit-adm2-eta}
% \end{equation}
% where $\mu_c$ is the underlying prevalence or rate in cluster $c$, which we typically assume to be the same for clusters from the same area. $\bm{x}_{j[c]}$ are covariates for the subarea containing cluster $c$, and $b_{j[c]}$ is a subarea-level random effect (as in the FH model). 

The mean structure is specified through a link function:
\begin{equation}
g(\mu_c) = g(\mu_{c[j]}) =\alpha + \bm{x}_{j}^\top \bm{\beta} + b_{j},
\label{eq:unit-adm2-eta}
\end{equation}
where $\mu_c$ is the underlying prevalence or rate in cluster $c$, which we typically assume to be the same for clusters from the same area. Let $c[j]$ denote the subarea $j$ in which cluster $c$ resides. We then have $\bm{x}_{j}$ as the covariates and $b_{j}$ as a subarea-level random effect (as in the FH model) in subarea $j$. 

A limitation of unit-level models is that the survey weights are not directly incorporated, which means that the sampling design is not explicitly accounted for. This becomes especially problematic in stratified designs, in which ignoring strata such as urban/rural status can yield biased estimates when the stratification variable is associated with the outcome \textit{and} there is over- or under-sampling of strata~\citep{wu2024modelling}. A viable solution is to include such stratification variables directly in the mean structure to adjust for their effect and then aggregate strata-specific estimates to overall estimates. %Such a task can be resolved in the following formulation. 

\subsubsection{MRP}
\label{sec:MRP-base}
MRP provides a principled framework for combining model-based estimation with survey design information and demographic covariates. For example, suppose we aim to conduct stratification by demographic groups $a \in \mathcal{A}$ (e.g., age, sex, or urban/rural categories).  %By incorporating demographic heterogeneity, MRP enables subarea estimates even when survey outcomes are observed only at coarser geographic levels. 
We can specify the hierarchical models in terms of group-specific likelihoods for group-specific means, both for the FH and unit-level approaches. Outcomes are modeled within poststratification cells defined by stratifying variables such as urban/rural status and demographic categories.

For the FH model, we use
\begin{align*} \hat \lambda_{j,a}^w \mid \lambda_j &\sim \mathcal{N}(\lambda_{j,a}, V_{j,a}), \\ \lambda_{j,a} &= \alpha + \bm{x}_j^\top \bm{\beta} + b_j + f(a), \end{align*}
with group-specific prevalence or rates obtained by transforming back from the modeling scale:
\[
\mu_{j,a} = g^{-1}(\lambda_{j,a}).
\]

For the unit-level model, we use
\[
Y_{c,a} \mid \mu_{c,a} \sim 
\begin{cases}
\text{Binomial}(n_{c,a}, \mu_{c,a}), & \text{binary outcomes}, \\
\text{Poisson}(n_{c,a} \mu_{c,a}), & \text{count outcomes},
\end{cases}
\]
\[
g(\mu_{c,a}) = g(\mu_{c[j],a}) = \alpha + \bm{x}_{j}^\top \bm{\beta} + b_{j} + f(a).
\]

The stratification effects, denoted as $f(a)$, are included alongside covariates and spatial random effects. When grouping variables are few and categories are limited, fixed effects (and their interactions) can be modeled directly. However, as the number of groups grows, data within each cell become sparse. In such settings, hierarchical smoothing can be adopted to stabilize estimates.

In the poststratification stage, predictions are aggregated to areal-overall estimates by using external population counts, $N_{j,a}$, typically from census data, so that estimates reflect the true demographic composition:
\begin{equation}
\mu_j^{\text{MRP}} = \frac{1}{N_j} \sum_a N_{j,a} \, \mu_{j,a}, 
\qquad N_j = \sum_a N_{j,a}.
\label{eq:MRP-aggre}
\end{equation}

Thus, MRP extends either FH or unit-level models by explicitly adjusting for predictive demographic variables and ensuring that estimates align with population structure. This adjustment accounts for subarea variability arising from demographic heterogeneity and improves the precision of small-area estimates.  

\subsection{Areal Disaggregation}

Our proposed methods for areal disaggregation extend beyond the standard SAE framework introduced above. For example, the survey outcomes are geo-indexed only at a coarser level, but the analytic and policy objectives require estimates at a finer resolution. 

To be specific, suppose a survey record's data $\{y_i, n_i\}$ are geo-indexed at the coarser area $i$, and the estimation target is the prevalence or rate $\mu_{j}$ for subareas $j$ nested within $i$. The estimation of $\{\mu_{j}\}$ cannot be made without additional assumptions or information because the distribution of outcomes within each coarse area is not identifiable from aggregate data alone. 

Our proposed framework addresses this challenge by systematically incorporating auxiliary sources of information that can inform within-area variation. These sources can be grouped into three broad categories. First, spatial structure enables borrowing strength across neighboring subareas. By building the proximity matrix at a finer area level, we can extend the spatial patterns in coarser regions to a finer spatial level.  Second, covariates ranging from environmental to demographic indicators from external sources (e.g., satellite data, census data) help explain outcome variation within coarse areas. Third, MRP improves estimates by aggregating group-specific estimates with the true population demographic composition such that the heterogeneity in the demographics can inform the variation in the subareas---even if the group-specific estimates are the same across subareas.

The following subsections describe the statistical models that perform areal disaggregation by combining these complementary sources of information.

\subsubsection{FH Disaggregation}
\label{sec:FH-model-disaggre}

We first extend the area-level (FH) model for disaggregation. We start with survey-weighted direct estimates at coarse area $i$. Specifically, let $\hat{\lambda}^{w}_i$ denote the transformed direct estimate for area $i$, with design-based variance $V_i$. Following the setup for an FH model, we assume a Gaussian distribution:
\begin{equation*}
\hat{\lambda}^{w}_i \mid \lambda_i \sim \mathcal{N}(\lambda_i, V_i),
\label{eq:data_model}
\end{equation*}
where $\lambda_i$ is the latent prevalence or rate in area $i$. 

To link the area-level parameter $\lambda_i$ to a latent process defined at a fine spatial level,
we construct a population-weighted aggregation of subarea means:
\begin{equation}
\lambda_i = g(\mu_i) = g\left( \sum_{j : i[j] = i} \frac{N_j}{N_i} \mu_j \right),
\label{eq:FH-subarea-aggre}
\end{equation}

where $N_j$ is the population of subarea $j$, $N_i = \sum_{j : i[j] = i} N_{j}$ is the population for coarser area $i$, and $g(\cdot)$ is the link function.

The latent process is indexed at the subarea level. Specifically, for each subarea $j$, we have
\begin{equation*}
g(\mu_j) = \alpha + \bm{x}_j^\top \bm{\beta} + b_j,
\label{eq:latent}
\end{equation*}
where the setup for modeling parameters and smoothing structures mirrors the FH model discussed in Section \ref{sec:FH-model-admin2}. 

We defer the computational details and implementation aspects of the Bayesian hierarchical model to Section \ref{sec:model-implementation-inlabru}. In particular, the nonlinear aggregation in equation \eqref{eq:FH-subarea-aggre} requires special treatment.

% Standard Bayesian inference via MCMC will be computationally infeasible, we require tools for numerical integration for posterior inference. A popular choice, INLA builds on latent Gaussian model, however in our model, the mapping between the data likelihood parameter $p_i$ and the linear predictor $\eta_{ij}$ involves nonlinear transformation. The inlabru package can handle this nonlinearity. Another choice is stanTMB.

\subsubsection{Unit-Level Model Disaggregation}
\label{sec:unit-model-disaggre}

We extend the framework to the unit-level model, in which observations are recorded at clusters, $c$, nested within coarse areas, $i$. Let $Y_c$ denote the observed outcome in cluster $c$ with a cluster size of $n_{c}$. The objective is to link the cluster-level outcomes that are geo-indexed at the coarser level $i[c]$ to a latent process defined at finer subareas, $j$.

In this setting, outcomes within the same cluster are not independently drawn from a single shared mean but instead exhibit dependence. Because all individuals in cluster $c$ belong to the same subarea, their outcomes are more alike than if they were sampled independently from the area mean. This within-cluster similarity inflates variation in the aggregated outcomes, so standard Poisson or binomial likelihoods underestimate uncertainty. To address this, we adopt overdispersed marginal sampling models with the negative binomial for count outcomes and the beta-binomial for binary outcomes:
% \[
% Y_{i,c} \sim 
% \begin{cases}
% \text{NegBin}(n_{i,c}\mu_{i[c]}, \phi), & \text{count outcomes}, \\
% \text{BetaBinomial}(n_{i,c}, \mu_{i[c]}, \phi), & \text{binary outcomes},
% \end{cases}
% \]
% where $\phi$ is an overdispersion parameter capturing the excess variation induced by within-cluster dependence.

% Following the extension to the FH model, we link observed clusters at coarser area i to
% the latent structure indexed at fine subareas. The mean at area $i$ is expressed as a population-weighted average of subarea means,
% \begin{equation}
% \mu_{i[c]} =  \sum_{j : i[j] = i} \frac{N_j}{N_i}\,\mu_j,
% \label{eq:unit-subarea-aggre}
% \end{equation}
% where $N_j$ is the population of subarea $j$ and $N_i = \sum_{j : i[j] = i} N_j$ is the population of the coarser area.

\[
Y_{c} \sim 
\begin{cases}
\text{NegBin}(n_{c}\mu_{c}, \phi), & \text{count outcomes}, \\
\text{BetaBinomial}(n_{c}, \mu_{c}, \phi), & \text{binary outcomes},
\end{cases}
\]
where $\phi$ is an overdispersion parameter that captures excess variation induced by within-cluster dependence.

Following the extension of the FH model, we link observed clusters at coarser area $i$ to
the latent structure indexed at fine subareas. The mean model for clusters residing in area $i$ is expressed as a population-weighted average of subarea means:
\begin{equation*}
\mu_{c} = \mu_{i[c]} =  \sum_{j : i[j] = i} \frac{N_j}{N_i}\,\mu_j,
\label{eq:unit-subarea-aggre}
\end{equation*}
where $N_j$ is the population of subarea $j$, and $N_i = \sum_{j : i[j] = i} N_j$ is the population of the coarser area.

At the subarea level, the latent process is modeled through a linear latent field:
\begin{equation}
g(\mu_j) = \alpha + \bm{x}_j^\top \bm{\beta} + b_j,
\label{eq:unit-latent-disaggre}
\end{equation}
where $\bm{x}_j$ are covariates, $b_j$ are random effects, and $g(\cdot)$ is the link function. The random effects $b_j$ capture both structured and unstructured spatial variation.

% The use of Negative Binomial and Beta-Binomial distributions can alternatively be motivated from a mixture model perspective. In this view, cluster totals arise as mixtures of Poisson or Binomial distributions with mixture priors on their parameters. This interpretation highlights the role of overdispersion as an approximation to the variation generated by heterogeneous subarea-level means.

\subsubsection{Extension to Incorporate MRP}
\label{sec:disaggre-MRP}

With the above setup for unit-level and FH disaggregation models, we can naturally incorporate MRP. This framework allows demographic heterogeneity to inform subarea estimates, even when outcomes are observed only at coarser geographic levels. 

Let the stratification groups be $a \in \mathcal{A}$. Similar to the MRP from Section \ref{sec:MRP-base}, we define the data model and mean structure by groups.

For the FH model, we set the hierarchal structure as,
\begin{equation}
\begin{aligned}
\hat \lambda_{j,a}^w \mid \lambda_{j,a} &\sim \mathcal{N}(\lambda_{j,a}, V_{j,a}), \\
\lambda_{i,a} &= g\!\left( \sum_{j : i[j] = i} \frac{N_{j,a}}{N_{i,a}} \, \mu_{j,a} \right),\\
g(\mu_{j,a}) &= \alpha + \bm{x}_j^\top \bm{\beta} + b_j + f(a)
\end{aligned}
\label{eq:FH-model-disaggre-MRP}
\end{equation}
where $N_{j,a}$ is the population of group $a$ in subarea $j$, $N_{i,a} = \sum_{j : i[j] = i} N_{j,a}$ represents the population in coarser area $i$.

Similarly, for the unit-level model, we have

\begin{equation}
\begin{aligned}
Y_{c,a} \mid \mu_{c,a} &\sim 
\begin{cases}
\text{Binomial}(n_{c,a}, \mu_{c,a}), & \text{binary outcomes}, \\
\text{Poisson}(n_{c,a} \mu_{c,a}), & \text{count outcomes},
\end{cases} \\
\mu_{c,a} &= \mu_{c[i],a} = \sum_{j : i[j] = i} \frac{N_{j,a}}{N_{i,a}} \, \mu_{j,a}, \\
g(\mu_{c,a}) &= \alpha + \bm{x}_{j[c]}^\top \bm{\beta} + b_{j[c]} + f(a).
\end{aligned}
\label{eq:unit-model-disaggre-MRP}
\end{equation}

Everything follows the standard MRP detailed in Section \ref{sec:MRP-base}. We can aggregate back to the overall estimates by using the same formulation detailed in Eq.~\eqref{eq:MRP-aggre} because the mean structures are all defined at fine areas $j$.

% This extension enables demographic composition to shape the distribution of outcomes across subareas. Hierarchical modeling of $f(a)$ allows borrowing strength across groups, while poststratification ensures that the resulting estimates align with the actual population structure. 

\subsection{Model Implementation and Nonlinearity}
\label{sec:model-implementation-inlabru}
The integrated nested Laplace approximation (INLA) method implemented via the \texttt{R-INLA} package (version 25.06.22) is widely used for Bayesian small-area estimation because it provides fast and accurate approximate inference for latent Gaussian models (LGMs) \citep{bakka2018spatial,rue2009approximate,lindgren2015bayesian,jontwoculture2025}. For spatial hierarchical models, \texttt{R-INLA} exploits the sparse precision structure of Gaussian Markov random fields to achieve substantial computational efficiency and incorporates built-in spatial smoothing priors such as the BYM2 prior \citep{riebler2016an}, which we adopt in our analysis. These features make \texttt{R-INLA} a natural choice for implementing our modeling framework.

The baseline INLA method assumes a linear predictor of the form $\bm{\eta} = \bm{A}\bm{u}$,
where $\bm{u}$ collects the latent Gaussian components, including fixed effects and both spatial and non-spatial random effects (e.g., the right-hand side of Eq.~\eqref{eq:FH-adm2-eta}). Each observation $y_k$ depends on $\eta_k$ through a possibly nonlinear link $h(\cdot)$ in the sampling model. In the FH model, $h(\cdot)$ is the identity link because the transformation of the outcome is applied before the sampling model, whereas for unit-level models, $h(\cdot) = g^{-1}(\cdot)$. Thus, the only nonlinearity permitted is the link from $\eta_k$ to the location parameter; the mapping from $\bm{u}$ to $\bm{\eta}$ must be strictly linear \citep{rue2009approximate}. Models geo-indexed at fine-level subareas conform to this constraint and can be implemented directly in \texttt{R-INLA}, as specified in Eqs.~\eqref{eq:FH-adm2-eta} and \eqref{eq:unit-adm2-eta}.

In contrast, our disaggregation models violate this requirement because aggregation introduces nonlinearity. The predictors linked to the sampling model’s location parameter become nonlinear combinations of the latent Gaussian components.

To demonstrate this, we start from the latent field in Eq.~\eqref{eq:unit-latent-disaggre}, and the subarea mean is
\[
  \mu_j = g^{-1}\!\left(\alpha + \bm{x}_j^\top \bm{\beta} + b_j\right) 
  = g^{-1}(\bm{u}_j).
\]
For both the FH and unit-level disaggregation models, the predictors take the form of
\begin{equation}
  \tilde{\eta}_i(\bm{u})
  \;=\;
  g\!\left(\sum_{j:i[j]=i} \frac{N_j}{N_i}\,\mu_j\right)
  \;=\;
  g\!\left(\sum_{j:i[j]=i}\frac{N_j}{N_i}\,
  g^{-1}(\bm{A}\bm{u}_j)\right),
  \label{eq:fh-nonlinear-predictor}
\end{equation}
where the nonlinear predictor $\tilde{\eta}_i(\bm{u})$ involves transformations of aggregated predictors\footnote{A discussion on why an apparently simpler formulation for unit-level models is not compatible with the \texttt{inlabru} framework is provided in Section~\ref{supp:identity-vs-log-link} of the supplemental material.} (e.g., log of weighted sums of exponentials, logit of weighted sums of expits) and cannot be expressed in linear form $\tilde{\bm{A}}\bm{u}$.

% where the nonlinear predictor $\tilde{\eta}_i(\bm{u})$ involves transformations of aggregated predictors (e.g., log of weighted sums of exponentials, logit of weighted sums of expits) and cannot be expressed in linear form $\tilde{\bm{A}}\bm{u}$. In the unit-level, one might be tempoted to consider a simplified formulation for the predictor through an identity link, but it is not compatible with the upcoming inlabru formulation. A detailed discussion of this point is provided in Supplement~X.

To overcome this challenge, we leverage the \texttt{inlabru} package (version 2.13.0.9011) \citep{lindgren2024inlabru,bachl2019inlabru,inlabruRpackage}, which extends \texttt{R-INLA} to allow latent predictors that are nonlinear functions of the latent Gaussian field. In the \texttt{inlabru} formulation, we express the stacked vector of nonlinear predictors $\tilde{\bm{\eta}}(\bm{u})$ as a deterministic function of $\bm{u}$, as in Eq.~\eqref{eq:fh-nonlinear-predictor}. Choosing a linearization point $\bm{u}_0$, we take a first-order Taylor approximation:
\begin{equation*}
  \bar{\bm{\eta}}(\bm{u})
  \;=\;
  \tilde{\bm{\eta}}(\bm{u}_0)
  + \bm{B}\,(\bm{u}-\bm{u}_0)
  \;=\;
  \bm{B}\bm{u} 
  + \bigl[\tilde{\bm{\eta}}(\bm{u}_0)-\bm{B}\bm{u}_0\bigr],
\end{equation*}
where $\bm{B}$ is the derivative matrix of the nonlinear predictor evaluated at $\bm{u}_0$. This yields a linearized version of the original nonlinear predictor with an offset depending on $\bm{u}_0$.

The exact marginal sampling model is defined by
\[
  \bm{y} \mid \bm{u}, \bm{\theta}
  \;\sim\;
  p\!\left(\bm{y} \,\middle|\, \tilde{\bm{\eta}}(\bm{u}),\, \bm{\theta}\right),
\]
where $\bm{\theta}$ is the hyperparameters. We approximate this by replacing the nonlinear predictor with its linearization, which gives us
\begin{align*}
  \bar{p}(\bm{y} \mid \bm{u}, \bm{\theta})
  \;=\;
  p\!\left(\bm{y} \,\middle|\, \bar{\bm{\eta}}(\bm{u}), \bm{\theta}\right)
  \;\approx\;
  p\!\left(\bm{y} \,\middle|\, \tilde{\bm{\eta}}(\bm{u}), \bm{\theta}\right)
  \;=\;
  \tilde{p}(\bm{y} \mid \bm{u}, \bm{\theta}).
\end{align*}

Through this linearization, the nonlinear observation model can be approximated by a standard LGM that can be fitted with INLA, thereby allowing access to the full suite of INLA-based inference and posterior summaries.

A key step in the \texttt{inlabru} method is the choice of a suitable linearization point $\bm{u}_0$. This is achieved through a fixed point iteration procedure, which updates the linearization point so that it is consistent with the conditional posterior mode of the linearized model. Starting from an initial point, the algorithm alternates between applying the INLA method to update the hyperparameters and computing the conditional posterior mode of the latent field, refining $\bm{u}_0$ until convergence is achieved. Full details of this iterative scheme, including the update rules and convergence criteria, are provided by \citet{lindgren2024inlabru}.

% \paragraph{Iterative scheme.}
% \texttt{inlabru} implements a fixed-point iteration:
% (i) start from $\mathbf u_0$; 
% (ii) build the linearised predictor at $\mathbf u_0$ and fit the linearised model with INLA to obtain $(\hat{\bm\theta}, \hat{\mathbf u})$; 
% (iii) update the linearisation point by a line search 
% $\mathbf u_0 \leftarrow (1-\alpha)\,\mathbf u_0 + \alpha\,\hat{\mathbf u}$ 
% that minimises a norm of the difference between $\tilde\eta$ and its linearisation; 
% (iv) repeat until convergence (e.g., small relative change and $\alpha \approx 1$).
% This yields an accurate approximate posterior for models whose predictor is genuinely nonlinear in $\mathbf u$.

% In order to calculate the derivatives matrix required for the linearisation step, inlabru uses
% numerical derivatives rather than requiring the user to directly specify the derivatives of the
% predictor formula. 

\section{Simulation}
\label{sec:simulation}

\subsection{Overview}

We designed a simulation study to assess the performance of spatial disaggregation methods for estimating the GFR in Kenya. In our data-generating process, individual-level fertility outcomes are first simulated and geo-referenced at the Admin-2 level (constituency), which serves as the ground truth. We then replicate the sampling procedure of a DHS to draw survey samples from this synthetic population. For model fitting, we impose a data restriction: only the geo-indexed Admin-1 (county) level data are made available, and all Admin-2 identifiers are removed. The proposed disaggregation method is then applied to infer Admin-2 level GFR estimates from the Admin-1 aggregates. Finally, these estimates are compared to the known Admin-2 truths to quantify the accuracy and reliability of the disaggregation approach.

Throughout, we use the conventional period definition of the GFR as births per 1,000 woman-years of exposure among women of reproductive age (15--44 years old): 
\[
\mathrm{GFR} 
= 1000 \times 
\frac{\displaystyle \sum_{a \in A} B_a}
{\displaystyle \sum_{a \in A} E_a / 12},
\]
where A = $\{15\text{--}19, \; 20\text{--}24, ..., \; 40\text{--}44\}$ represents age groups, $B_a$ is the total number of live births in age group $a$ during the reference period, and $E_a$ is the total woman-months of exposure in the age group $a$. Division by $12$ converts months of exposure to woman-years, and the factor of $1{,}000$ scales the rate to births per $1{,}000$ woman-years.
In our simulations, the reference window is the 60 months preceding the survey year, and we target $T^\star=48$ months of mean exposure per woman.

%Administrative boundaries follow IEBC (citation needed). 
Based on Kenya's administrative boundaries, in the analysis, we treat Admin-1 as the 47 counties and Admin-2 as the 290 constituencies. The simulated sample size and stratification emulate the 2022 Kenya DHS, which is based on the 2019 census as a sampling frame.

\subsection{Data Sources, Construction of Master Frame, and Sampling}

We align administrative boundary shapefiles with the 2019 Kenya census to construct sociodemographic marginals for each Admin-2 unit. The key variables include the proportion of urban population, the proportion of women with secondary or higher education, and additional area-level covariates such as average household size and mobile phone access.

Population counts are drawn from WorldPop data \citep{tatem2017worldpop}. Gridded, age-specific population estimates are aggregated to the Admin-2 level to serve as the base population. Additional raster covariates, such as nighttime lights, health facility access, and vegetation indices are calculated as population-weighted averages at the Admin-2 level.

To generate the master frame, we adopt the sampling design of the Kenya DHS 2022. A synthetic population is created by expanding Admin-2~$\times$~age~$\times$~urbanicity~$\times$~education cells into individual records. Enumeration areas, or clusters, are then allocated to strata in proportion to the female population.

Finally, we mimic the DHS two-stage sampling strategy. Clusters are selected within strata by using a probability proportional to the size of the population, after which 20 women per cluster are sampled at random. The resulting inclusion probabilities define the base sampling weights to ensure that the simulation closely replicates the DHS design. Additional details are provided in Section~\ref{sec:supp-simulation-setup} of the supplemental material.

\subsection{Outcome Generation and Simulation Scenarios}

We aim to construct several simulation scenarios that represent potential sources of variability in subnational outcome estimates. The goal is to start with a simple scenario and then progressively include additional factors that better approximate real-world data. Specifically, we consider four scenarios, summarized in Table \ref{tab:four_simulation_scenarios}, that differ by their inclusion of observed covariates, unobserved covariates, area-level independent random effects, and model misspecification. 

In the first scenario, the observed covariates can be naturally captured through the disaggregation model. In the second scenario, we simulate a case in which unobserved covariates, which often exhibit spatial dependence, can be partly accounted for through spatial smoothing that captures spatially correlated random effects at the Admin-2 level. In the third scenario, we introduce independent random effects at the Admin-2 level that cannot be identified from data geo-indexed at the Admin-1 level and therefore cannot be recovered. In the last scenario, we introduce model misspecification by assigning different coefficient values for urban and rural strata, in contrast to the global effect estimates assumed in the other models. Although the introduction of an interaction term partially reflects reality, we note that the actual complexities are likely greater than those captured in even our most complex scenario.

To formalize the outcome generation process, let $E_{c,k}$ denote the exposure during the reference period for individual $k$ in cluster $c$. The individual-level birth process is generated independently as
\[
Y_k \sim \mathrm{Poisson}(E_{c,k}\,\mu_{c,k}),
\]
where $\mu_{c,k}$ is the individual fertility rate.

\begin{table}[ht]
\centering
\caption{Sources of spatial heterogeneity under four scenarios and the expected ability of disaggregation models to capture them.}
\renewcommand{\arraystretch}{1.4}
\begin{tabular}{|l|c|c|c|c|c|}
\hline
\textbf{Variation sources} & \textbf{Scenario 1} & \textbf{Scenario 2} & \textbf{Scenario 3} & 
\textbf{Scenario 4} & \textbf{Recoverable} \\ \hline

\makecell[l]{Observed covariates} &
\cmark &
\cmark &
\cmark &
\cmark &
\makecell[l]{\cmark \\ Explicitly captured \\ by fixed effects} \\ \hline

\makecell[l]{Unobserved covariates \\ (spatially correlated \\ Admin-2 effects)} &
\xmark &
\cmark &
\cmark &
\cmark &
\makecell[l]{\opencircle \\ Partially captured \\ by spatial effects} \\ \hline

\makecell[l]{IID Admin-2 \\ random effects} &
\xmark &
\xmark &
\cmark &
\cmark &
\makecell[l]{\xmark \\ Not recoverable, \\ purely unstructured} \\ \hline

\makecell[l]{Varying \\ coefficients} &
\xmark &
\xmark &
\xmark &
\cmark &
\makecell[l]{\xmark \\ Leads to model \\ misspecification} \\ \hline

\end{tabular}
\label{tab:four_simulation_scenarios}
\end{table}

\paragraph{Scenario 1:} In the first scenario, we include only observed covariates. The log-scale mean structure for individual $k$ in cluster $c$ located in Admin-2 area $j$ is specified as
\begin{align*}
\log \mu^{s_1}_{c,k} &=  \alpha_{\mathrm{rural}}\,I(c \in \text{rural})
    + \alpha_{\mathrm{urban}}\,I(c \in \text{urban}) \\
    &\quad + \delta_{\mathrm{age}(k)} + \delta_{\mathrm{educ}(k)} \\
    &\quad + \bm{\beta}\bm{X}_{j[k]} + e_k + e_c.
\end{align*}
This structure incorporates settlement-type intercepts, individual-level covariates (age and education), area-level covariates (e.g., nighttime lights, travel time to nearest health facility, household size), and random effects at the individual and cluster levels. Full parameter specifications for this and later scenarios are available in Section~\ref{sec:supp-scenario} of the supplementary material.

\medskip
\paragraph{Scenario 2:} The second scenario extends this mean structure by including additional unobserved covariates at the area level. Specifically,
\[
\log(\mu^{s_2}_{c,k}) = \log(\mu^{s_1}_{c,k}) + \bm{\beta}_{\mathrm{unobs}}\bm{X}^{\mathrm{unobs}}_{j[k]},
\]
where $\bm{X}^{\mathrm{unobs}}_{j[k]}$ introduces further variation through two unobserved covariates: the vegetation index and female mobile phone usage, both of which are standardized. These covariates are excluded from the modeling process, but disaggregation models are expected to partially capture these variations through spatially correlated Admin-2 effects.

\medskip
\paragraph{Scenario 3:} In the third scenario, we allow for additional unexplained heterogeneity by introducing an independent Admin-2 level random effect:
\[
\log(\mu^{s_3}_{c,k}) = \log(\mu^{s_2}_{c,k}) + e_j, \qquad e_j \sim \mathcal{N}(0, \sigma^2_j).
\]
This captures unstructured residual variation that the disaggregation model cannot explain.

\medskip
\paragraph{Scenario 4:} The fourth scenario introduces model misspecification by allowing coefficients to differ between urban and rural areas within the same Admin-2 unit. This specification permits heterogeneity across urban and rural populations:
\[
\log(\mu^{s_4}_{c,k}) = \alpha^{(s)} + \delta^{(s)}_{\mathrm{age}(k)} + \delta^{(s)}_{\mathrm{educ}(k)} + \bm{\beta}^{(s)}\bm{X}_{j[k]} + e_k + e_c,
\]
where $s \in \{\mathrm{urban}, \mathrm{rural}\}$. 

\medskip
Although Scenario~4 offers a closer approximation to reality than the earlier scenarios, it might still underrepresent the complexity of actual data generating processes.

\medskip
In Scenarios 1 and 2, the synthetic population is generated once and reused across repetitions. In contrast, for Scenarios 3 and 4, in which the independent Admin-2 level effects can substantially influence the results, a new synthetic population is generated for each repetition. For each scenario, we draw repeated samples 500 times to yield 500 distinct datasets. Model fitting and performance metrics are then calculated based on the average across these 500 datasets.

\subsection{Models Considered and Comparison Metrics}

\begin{table}[ht]
\centering
\caption{Summary of models compared in the simulation study.}
\renewcommand{\arraystretch}{1.3}
\begin{tabular}{|p{3cm}|p{7cm}|p{2.5cm}|p{3cm}|}
\hline
\textbf{Model} & \textbf{Configuration} & \makecell[l]{\textbf{Geo-indexing}\\ \textbf{level}}  &\textbf{Specification} \\
\hline
FH disagg & Fay--Herriot disaggregation model; no MRP. & Admin-1 & \S\ref{sec:FH-model-disaggre}\\
\hline
FH-MRP disagg & Fay--Herriot disaggregation model; MRP by age and education (no urban/rural). & Admin-1& \S\ref{sec:disaggre-MRP},  Eq.\eqref{eq:FH-model-disaggre-MRP} \\
\hline
Unit disagg & Unit-level disaggregation model; stratified by urban/rural. & Admin-1 & \S\ref{sec:unit-model-disaggre}\\
\hline
Unit-MRP disagg & Unit-level disaggregation model; MRP by age, education, and urban/rural. & Admin-1& \S\ref{sec:disaggre-MRP},  Eq.\eqref{eq:unit-model-disaggre-MRP}\\
\hline
\makecell[l]{Direct Admin-2\\(benchmark)} & Survey-weighted direct estimates at Admin-2. & Admin-2 & S\ref{sec:direct-model-admin2} \\
\hline
\makecell[l]{FH Admin-2 \\(benchmark)}& Fay--Herriot model fitted directly at Admin-2. With observed covariates. & Admin-2& \S\ref{sec:FH-model-admin2} \\
\hline
\end{tabular}
\label{tab:models}
\end{table}

We compare four disaggregation models defined at the Admin-1 level with two benchmark approaches based on Admin-2 data (Table \ref{tab:models}). The disaggregation models are either FH or unit-level, and each has two variants: with and without MRP. In all cases, the latent structure is defined at the Admin-2 level. In disaggregation models, latent fields are aggregated to the Admin-1 scale by using the \texttt{inlabru} framework so that predictions can be linked to the observed data.

The FH disaggregation models use survey-weighted direct estimates at the Admin-1 level and are modeled on the logit scale. The FH-MRP variant extends this by applying MRP across demographic groups, although the urban/rural dimension is excluded because of sparse data and because survey weights already incorporate this stratification. Unit-level models are fitted to cluster-level outcomes, but the non-MRP specification still stratifies by urban/rural to mitigate the bias described earlier. The unit-MRP model also applies MRP by age, education, and urban/rural.

For benchmarking, we include two standard approaches that use data geo-indexed at Admin-2: direct survey-weighted estimates and an FH model fitted directly at Admin-2.

All models incorporate the same three observed covariates, and all unobserved covariates are excluded from model fitting. In the non-MRP variants, education is added as an area-level covariate and measured as the proportion of women with at least secondary education in each Admin-2 region.

We define a set of metrics to systematically evaluate the performance of the model in simulation scenarios. In the context of spatial disaggregation models, our primary interest lies in assessing the ability of the models to capture variation within Admin-1 regions rather than focusing solely on global or cross-regional variation. Specifically, we aim to determine how well the models leverage covariate information, spatial structures, and demographic composition to capture the relative ranking and heterogeneity of Admin-2 areas within the same Admin-1 area. 

Thus, we first focus on two metrics within-Admin-1: a regression-based \(R^2\) and a correlation-based measure. We denote the true GFR for Admin-2 area \(j\) as \(\mu_j\) and the corresponding model prediction as \(\hat{\mu}_j\). The true GFR for each Admin-1 group is $\bar{\mu}_i = \frac{1}{n_i}\sum_{j: i[j]=i} \mu_j$.
\vspace{0.3em}

\paragraph{Within-Admin-1 regression $R^2$:} For each Admin-1 region, we calculate the sum of squared errors (SSE) and the within-group total sum of squares ($\mathrm{SST}_{\text{within}}$):  
\[
\mathrm{SSE} \;=\; \sum_{i \in \mathcal{I}} \sum_{j: i[j]=i} \bigl(\mu_j - \hat{\mu}_j \bigr)^2, 
\qquad 
\mathrm{SST}_{\text{within}} \;=\; \sum_{i \in \mathcal{I}} \sum_{j: i[j]=i} \bigl(\mu_j - \bar{\mu}_i \bigr)^2.
\]  

The within-Admin-1 regression $R^2$ is then calculated as  
\[
R^2_{\text{within-reg}} \;=\; 1 - \frac{\mathrm{SSE}}{\mathrm{SST}_{\text{within}}}.
\]  
This measure quantifies the proportion of within-Admin-1 variance in the true outcomes explained by the model predictions rather than the total variance across all Admin-2.

\vspace{0.3em}

\paragraph{Within-Admin-1 Pearson correlation:} We use Pearson correlation to assess the rank of Admin-2 predictions. Within each Admin-1 area between the true and predicted GFR, we have  
\[
r_i \;=\; \mathrm{corr}\!\bigl(\mu_j, \hat{\mu}_j \;|\; i[j]=i \bigr).
\]  

The overall metric is obtained by averaging these correlations across all Admin-1 areas followed by squaring the mean correlation:  
\[
r_{\text{within-corr}}= \bar{r} \;=\; \frac{1}{|\mathcal{I}|} \sum_{i \in \mathcal{I}} r_i\text{.}
\]  
This correlation-based measure emphasizes the rank-order accuracy of predictions within Admin-1 regions.

\vspace{0.3em}

\paragraph{Uncertainty metrics:} We also evaluate the quality of uncertainty quantification provided by the models. Three metrics are considered: the average width of predictive intervals; the frequentist coverage of these intervals; and the interval score (IS), as defined by \cite{gneiting2007strictly}. 

Let \((l_{j}, u_{j})\) denote uncertainty intervals from the posterior predictive distribution of \(\hat{\mu}_{j}\). Specifically, the lower and upper bounds correspond to the \(\alpha/2\) and \(1 - \alpha/2\) quantiles, respectively. Interval quality is assessed using the IS by \citet{gneiting2007strictly}, which penalizes both overly wide intervals and those that fail to include the observed value. For a set of intervals \(\{(l_{ij}, u_{ij})\}\), the score is given by
\[
\mathrm{IS}_\alpha \;=\; \frac{1}{|\mathcal{J}|} \sum_{j \in \mathcal{J}}\Bigl[\, 
u_{j} - l_{j} 
+ \tfrac{\alpha}{2} (l_{j} - \mu_{j}) \,\mathbf{I}\!\bigl(l_{j} > \mu_{j} \bigr)
+ \tfrac{\alpha}{2} (\mu_{j}- u_{j}) \,\mathbf{I}\!\bigl(u_{j} < \mu_{j}\bigr)
\Bigr].
\]

This metric penalizes both overly wide intervals and those that fail to include the observed value. Lower scores correspond to better model performance.

\subsection{Simulation Results}
\begin{figure} [!ht]
    \centering
    \includegraphics[clip, trim=0cm 0cm 0cm 0cm, width=1\linewidth]{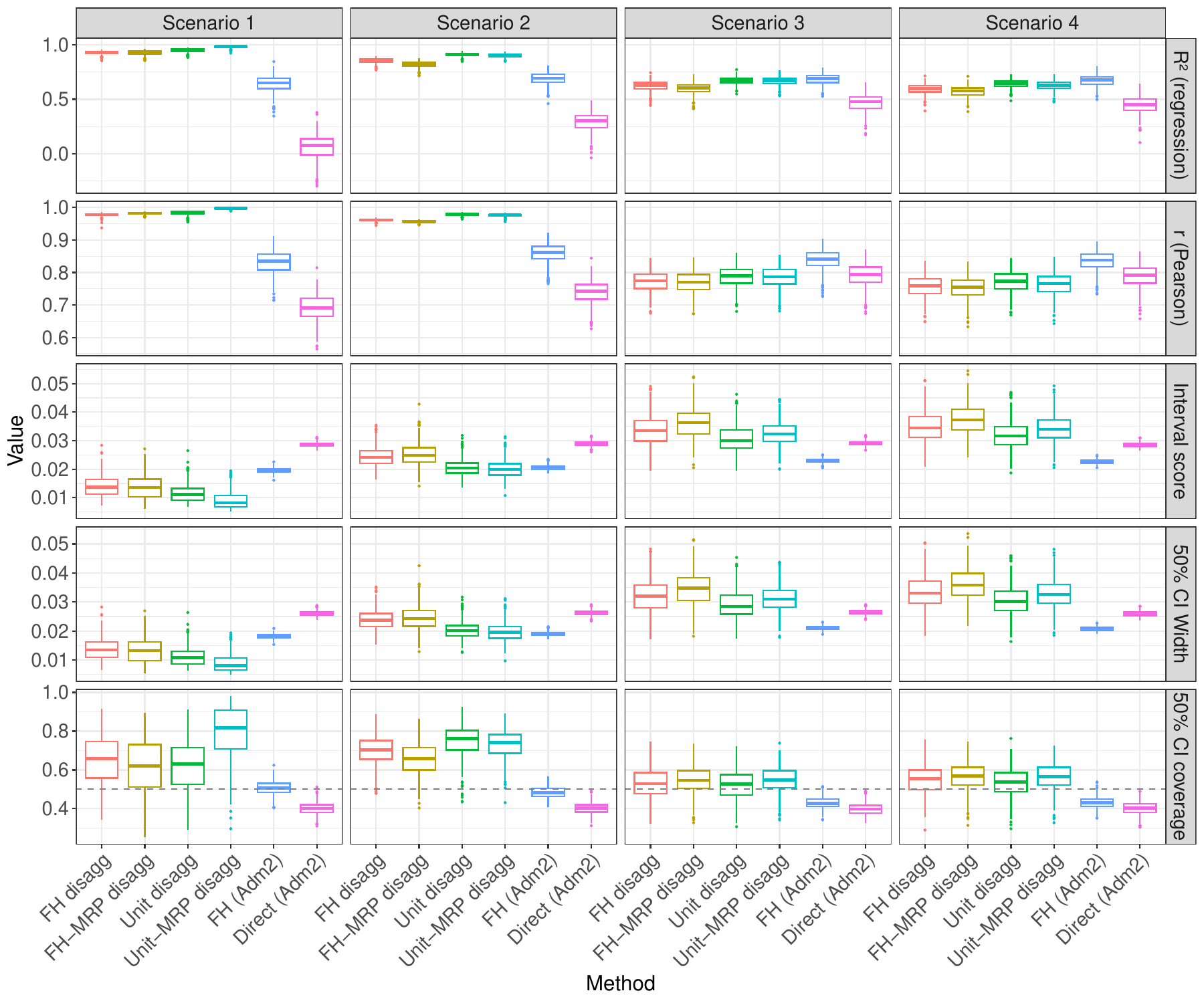}
    \caption{Comparison of disaggregation and benchmark models (FH and direct estimates at the Admin-2 level) across four simulation scenarios and evaluated with within-Admin-1 regression, $R^2$; Pearson correlation, $r$; the uncertainty IS; width; and coverage. Scenario 1 includes observed covariates and random effects at the individual and area levels. Scenario~2 expands scenario~1 with the introduction of two additional area-level measures that are assumed to be unobserved. Scenario~3 expands scenario~2 with the addition of IID Admin-2 level random effects. Scenario~4 is the most complicated scenario and introduces model misspecification by allowing the coefficients to differ between urban and rural areas within the same Admin-2 unit.}
    \label{fig:sim-res-box-measures}
\end{figure}

We present the performance of our proposed models and that of the benchmark models across the four predetermined scenarios. All models are fitted using the same set of observed covariates, with the exception of direct estimation, which does not incorporate covariates. Figure~\ref{fig:sim-res-box-measures} provides a summary of the models, and additional details are available in Section~\ref{sec:supp-simulation-res} of the supplemental material. In summary, our proposed models consist of the FH and unit-level disaggregation models, each with and without an MRP extension. These models are based on data geo-indexed at the Admin-1 level with Admin-2 identifiers removed. The benchmark models are FH models and direct estimations with data geo-indexed at the Admin-2 level.

Each experiment is repeated 500 times to mitigate the noise introduced by random sampling variation in the survey process. Metrics are then averaged over the 500 repetitions, and uncertainty bars are used to illustrate the variability in these metrics.

When comparing the disaggregation models with the benchmark models, a clear trend emerges. As the scenarios become increasingly complex and closer to real-world conditions, the performance of the disaggregation models decreases. In the last two scenarios, the FH Admin-2 model outperforms the disaggregation approaches. This pattern is consistent with the design of the scenarios. Moving from Scenario~1 to 4, we sequentially introduce sources of variation that the disaggregation models are less able to explain: observed covariates, unobserved covariates (representing spatially correlated effects), IID Admin-2 level random effects, and model specification error. The most significant turning point occurs with the introduction of IID Admin-2 random effects, which the benchmark models are able to capture owing to the sufficient resolution of the geo-indexed data. In contrast, the disaggregation models are unable to produce high-fidelity estimates when restricted to Admin-1 information.\clearpage

The weaker performance of the benchmark models in Scenarios 1 and 2 can be explained by their tendency to fit Admin-2 level independent random effects even though such effects were not present in those scenarios. This issue is particularly evident in direct estimation, whereas the FH Admin-2 model benefits from covariates and smoothing that partially stabilize the results. On the other hand, the exceptionally strong performance of the disaggregation models in Scenarios 1 and 2, with within-Admin-1's $R^2$ and Pearson's $r$ close to one, should be interpreted with caution. These scenarios were chosen to emphasize settings in which disaggregation models are expected to perform well and provide a clear baseline for evaluating their performance.

%These scenarios were deliberately constructed to favor disaggregation models and function more as a sanity check rather than a reflection of realistic applications.

Turning to the uncertainty measures, the disaggregation models show increased uncertainty in Scenarios 3 and 4 once the unidentifiable Admin-2 random effects are introduced. This is expected because these models are effectively extrapolating from Admin-1 to Admin-2. The tendency toward slight over-coverage is desirable when extrapolating because it reflects an appropriately conservative assessment of uncertainty. The IS values are higher in the last two scenarios for the disaggregation models versus the benchmark models. Again, this reflects the challenges of extrapolation. Overall, the variability across the 500 repetitions is also larger for the disaggregation models, thereby demonstrating their sensitivity to random sample variation. This is also expected because predictions in these models rely more heavily on the observed data.

Comparisons within the disaggregation models show that unit-level approaches perform slightly better than FH models across scenarios, although the margin is small. Considering the effect of the MRP extension, the MRP variants are slightly advantageous in Scenarios 1 and 2 but less effective in Scenarios 3 and 4. A plausible explanation is that the MRP variants attempt to attribute some of the unexplained variation that arises from IID Admin-2 random effects to demographic covariates, thereby introducing additional variability. For FH models, the MRP extension may also suffer from ecological bias because the models are fitted at the aggregate level, whereas the covariates used for post-stratification are at the individual level. This bias can be observed in the elevated values of the bias metric presented in Section~\ref{sec:supp-simulation-res}. For the unit-level models, however, post-stratification through MRP also helps counteract potential biases from informative sampling with respect to demographic variables because the unit-level models proposed cannot incorporate design weights.

In summary, the results show that the disaggregation models demonstrate reasonable performance. They can explain within-Admin-1 variation, as reflected in $R^2$ and Pearson's $r$, and their uncertainty quantification is not substantially worse than that of the benchmark models. At the same time, the simulations highlight the limitations of disaggregation models when faced with unidentifiable Admin-2 variation. Based on the findings from the simulation study, we focus on two models for subsequent real-data applications: the FH model without the MRP extension and the unit-level model with the MRP extension.

\begin{figure} [!ht]
    \centering
    \includegraphics[clip, trim=0cm 0.5cm 0cm 0cm, width=1\linewidth]{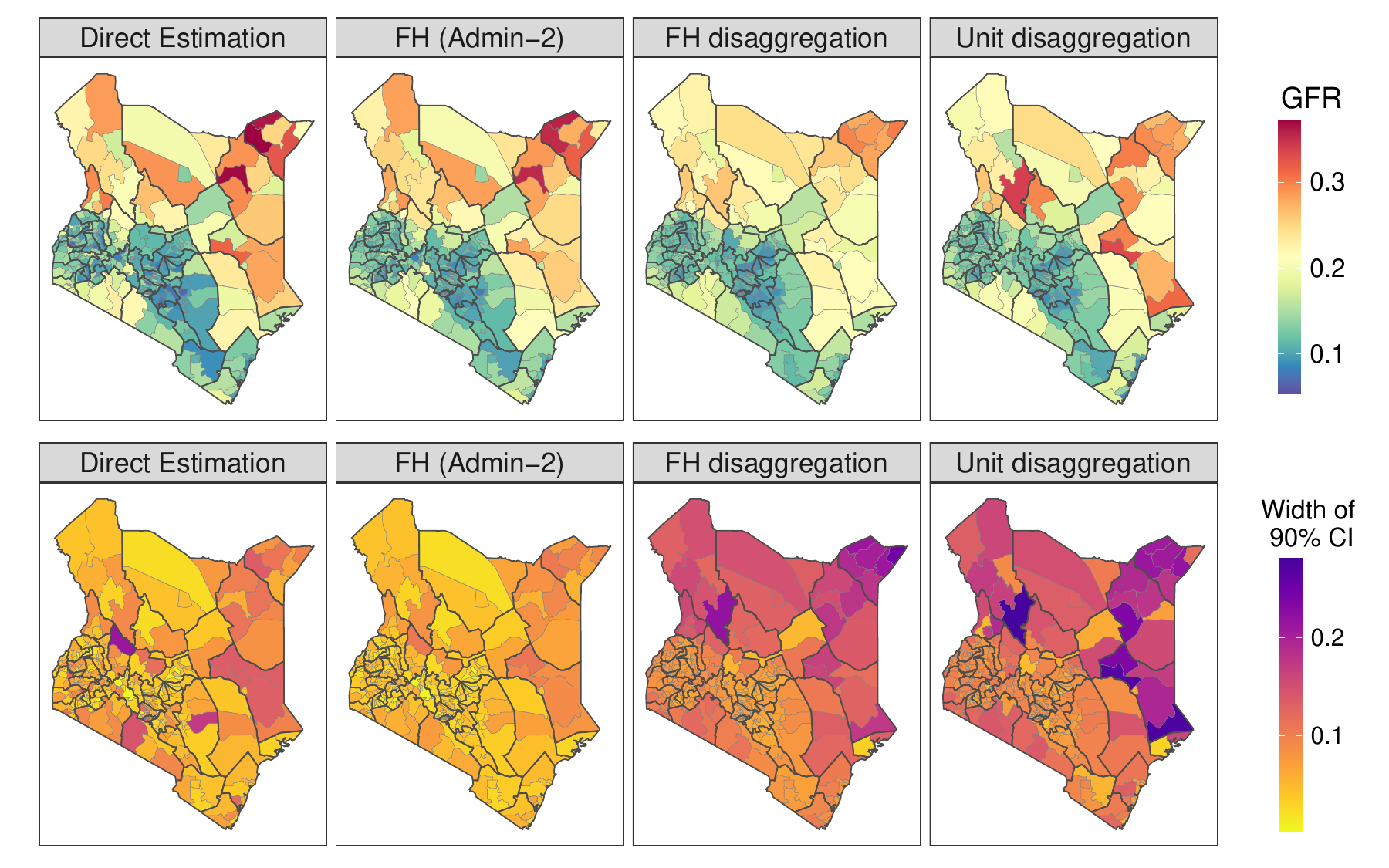}
    \caption{Admin-2 level estimates of GFR in Kenya based on DHS 2022 data and obtained from four models: direct estimation, FH at Admin-2, FH disaggregation, and unit-level disaggregation with MRP. The upper panel shows point estimates of GFR, and the bottom panel shows the width of the 90\% uncertainty intervals.}
    \label{fig:gfr-point-CI}
\end{figure}

\section{Case Study 1: GFRs in Kenya (DHS 2022)}
\label{sec:case-study-1}

Our simulation study was designed to mimic the data-generating and sampling process for the GFR in Kenya based on the 2022 DHS survey. In this case study, we apply the disaggregation models to the actual DHS 2022 data. Because the survey provides GPS coordinates for clusters, the data are geo-indexed at the Admin-2 level. We evaluate four models: two benchmark models that directly use Admin-2 indexing (direct estimation and the FH model with covariates) and two disaggregation models that assume only Admin-1 indexing is available (the FH model without MRP and the unit-level model with MRP). The covariates used are nighttime light intensity, travel time to the nearest health facility, average household size, mobile phone usage, and secondary school attainment rate.

The goal is to assess how disaggregation models perform when Admin-2 identifiers are masked, both relative to the benchmark models and in comparison with each other.

Figure \ref{fig:gfr-point-CI} shows that the disaggregation models produce Admin-2 level GFR estimates broadly consistent with the benchmark models. Larger differences arise in regions where the Admin-2 FH model estimates that the area-level random effects are greater, which is consistent with the simulation finding that the disaggregation models do not fully recover these effects by design (more details are provided in Section~\ref{sec:supp-case-study-1} of the supplemental material).

%The unit-level and FH disaggregation models disagree slightly in some Admin-2 regions. 

Taking the Admin-2 FH model as a benchmark, the average within-Admin-1 Pearson correlation $r$ between disaggregation and benchmark estimates is 0.52 for the FH model and 0.55 for the unit-level model, suggesting a reasonable level of agreement (correlations across all Admin-2 units exceed 0.9 at the global level). Nevertheless, there is still nontrivial uncertainty for the benchmark model estimates themselves, and these comparisons should be interpreted with caution.

Generally, uncertainty intervals are wider for the disaggregation models, which is consistent with the simulation study, given that disaggregation must extrapolate from Admin-1 to Admin-2. The benchmark models, in contrast, exhibit much narrower intervals because they directly exploit Admin-2 information, although such precision would be unattainable under Admin-1-only indexing.

Overall, the case study shows that disaggregation models can provide Admin-2 level estimates of GFR with reasonable accuracy when only Admin-1 geo-indexing is available. The trade-off is larger but appropriately conservative uncertainty intervals. 

\section{Case Study 2: Indicators from the 2021 KTUS}
\label{sec:case-study-2}

We now consider a case in which disaggregation methods are essential because the data are geo-referenced only at a coarse administrative level instead of the target level. Our application uses the 2021 KTUS, which was conducted as a module of the Kenya Continuous Household Survey. Similar to the DHS, the KTUS followed a two-stage stratified design that used the Kenya 2019 census as the master frame. Unlike the DHS, the KTUS does not provide GPS coordinates, so individual records are linked only to Admin-1 areas, making standard Admin-2 analysis impossible. Because policy and SDG monitoring require finer subnational detail, particularly for indicators such as the ones we consider below, disaggregation models are the only viable option.  

\subsection{Unpaid Care or Domestic Work}

\begin{figure} [!ht]
    \centering
    \includegraphics[clip, trim=0cm 0.5cm 0cm 0cm, width=1\linewidth]{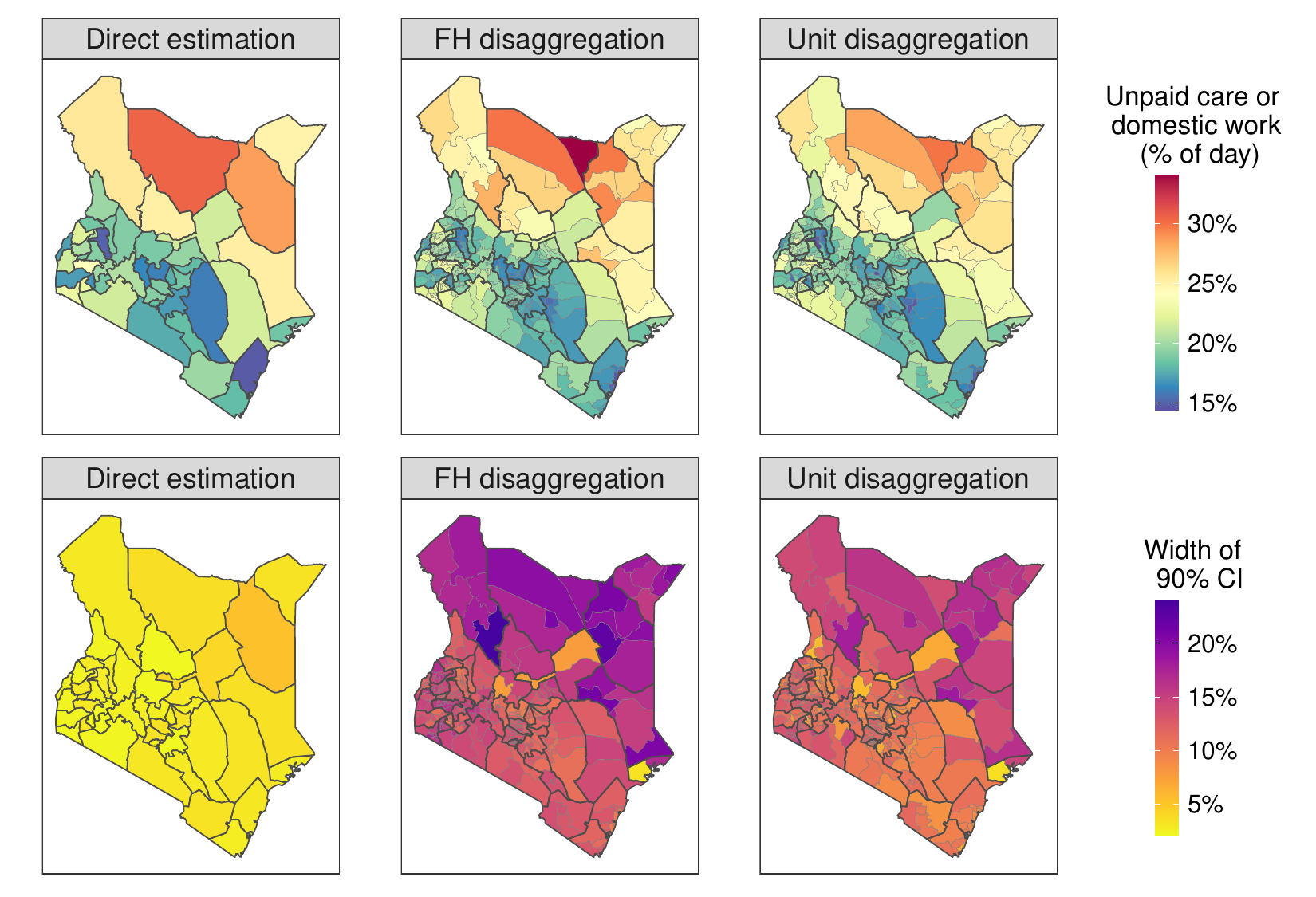}
    \caption{Estimated proportion of the day that women spend on unpaid care and domestic work from direct estimation at Admin-1, FH disaggregation, and unit-level disaggregation models. The maps show both point estimates and the widths of the associated 90\% uncertainty intervals.}
    \label{fig:female-unpaid-work}
\end{figure}

The first indicator considered is time spent on unpaid care or domestic work, which is central to SDG target 5.4.1. Respondents reported minutes spent on activities during the previous day, which we round to half-hour units. We apply negative binomial models with total half-hours for the day as exposure and compare three approaches: (1) direct estimation at Admin-1; (2) an FH disaggregation model without MRP; and (3) a unit-level disaggregation model with MRP using age group and urban/rural status for post-stratification. Covariates include nighttime light intensity, travel time to the nearest health facility, average household size, and high school attainment rate.

Figure~\ref{fig:female-unpaid-work} shows the estimated proportion of the day women spend on unpaid care and domestic work. The results range from about 15\% (3.5--4 hours) in some areas to more than 30\% (more than 7 hours) in others. Nairobi and Mombasa yield the lowest estimates, reflecting better access to services and formal employment, whereas the north and northeast show the highest estimates. Beyond these broad patterns, the disaggregation models further suggest substantial heterogeneity within Admin-1 units. Both disaggregation models highlight specific Admin-2 areas within Marsabit and Wajir near the Ethiopian border, where women devote more than 30\% of their time to unpaid care or domestic work. The two disaggregation models produce consistent estimates, with the unit-level model yielding smoother spatial patterns. Uncertainty intervals are wider for the disaggregation models than for direct Admin-1 estimates, reflecting the extrapolation from coarse to fine geographical levels. The male-specific results are shown in Figure~\ref{fig:male-unpaid-work} of the supplemental material. Overall, men spend less than 6\% of the day (under 1.5 hours) on unpaid care or domestic work.

\begin{figure} [!ht]
    \centering
    \includegraphics[clip, trim=0cm 0.5cm 0cm 0cm, width=1\linewidth]{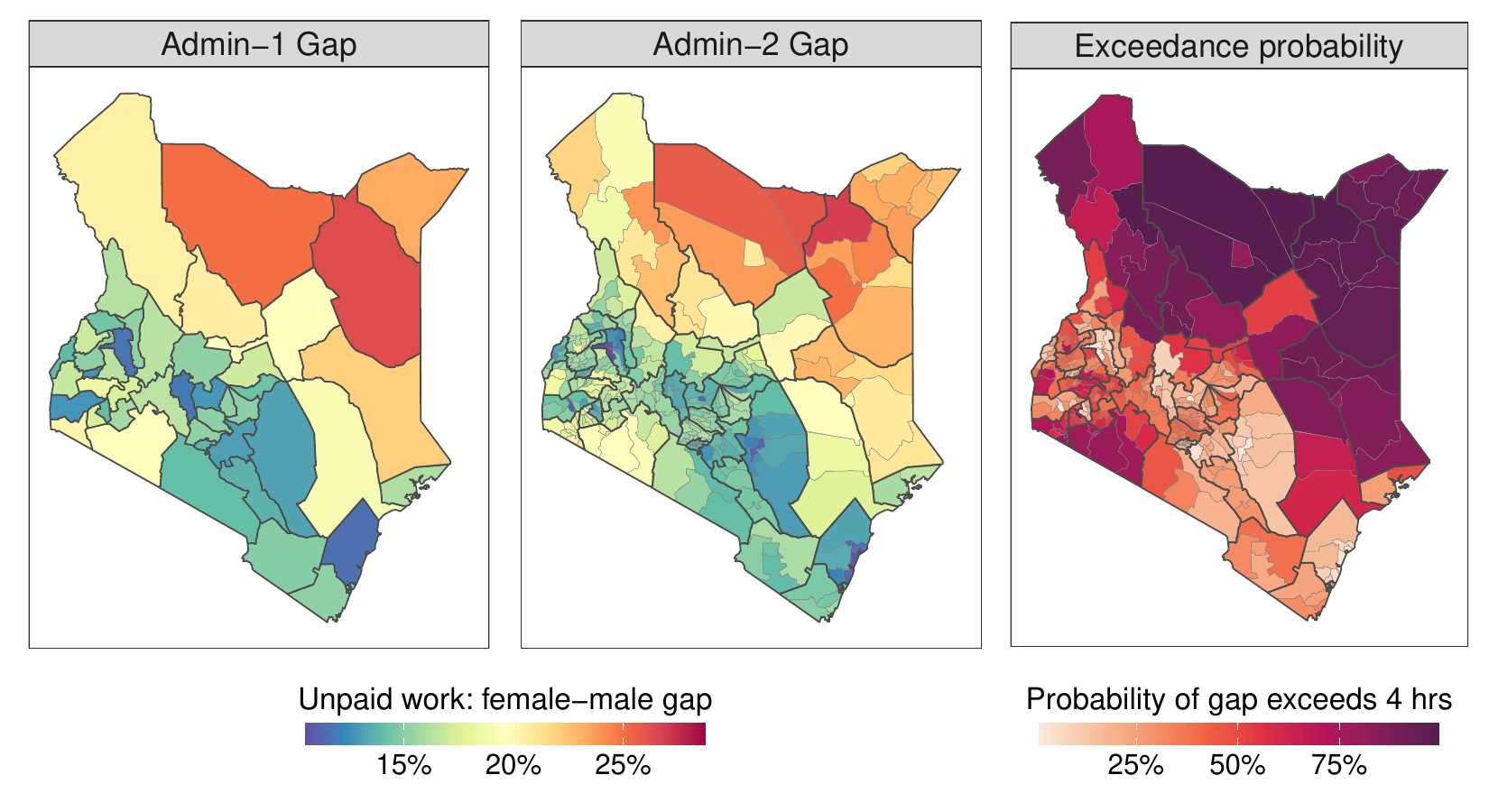}
    \caption{Maps of the female-male gap in unpaid care and domestic work and the associated probability that the difference exceeds four hours per day. Admin-1 estimates are based on direct estimates, and Admin-2 estimates are from a unit-level disaggregation model with MRP.}
    \label{fig:unpaid-work-gap}
\end{figure}

To examine gender gaps, we calculate the female-male difference in hours devoted to unpaid domestic and care work and map the probability that women spend more than four extra hours per day on such activities relative to men (equivalent to 16.7\% of the day). These exceedance probabilities are derived from posterior draws of the unit-level disaggregation model and are shown in Figure~\ref{fig:unpaid-work-gap}. The results reveal a clear spatial pattern: areas in the north and northeast consistently show probabilities above 75\%, indicating strong evidence of large gender gaps. In contrast, the central highlands and coastal areas (e.g., Mombasa and Kilifi) show much lower probabilities---in some cases below 25\%. Disaggregation reveals finer spatial patterns, with localized hot spots not visible in aggregated estimates. The male-female gap is particularly pronounced in these regions, underscoring the need for spatially targeted interventions for progression through SDG target 5.4.1.

\subsection{Mass Media Usage}
We now consider a binary indicator to demonstrate that the modeling framework also applies to binary outcomes rather than just rate outcomes (e.g., GFR). In this example, we focus on any mass media usage, which is defined as whether a respondent spent any time consuming mass media during the previous day. Mass media refers to communication channels that reach large audiences, such as radio, television, newspapers, and online news platforms. Mass media usage is a key indicator of information access and social participation, with strong links to public health messaging, civic engagement, and progress toward several SDG targets.  

The outcome is modeled by using the beta-binomial specification described in Section~\ref{sec:unit-model-disaggre}. As with the previous case, we compare three approaches: direct estimation at the Admin-1 level, FH disaggregation and unit-level disaggregation at the Admin-2 level. The covariates used are nighttime light intensity, high school attendance rate, mobile phone usage, and access to a health facility.

\begin{figure} [!ht]
    \centering
    \includegraphics[clip, trim=0cm 0cm 0cm 0cm, width=1\linewidth]{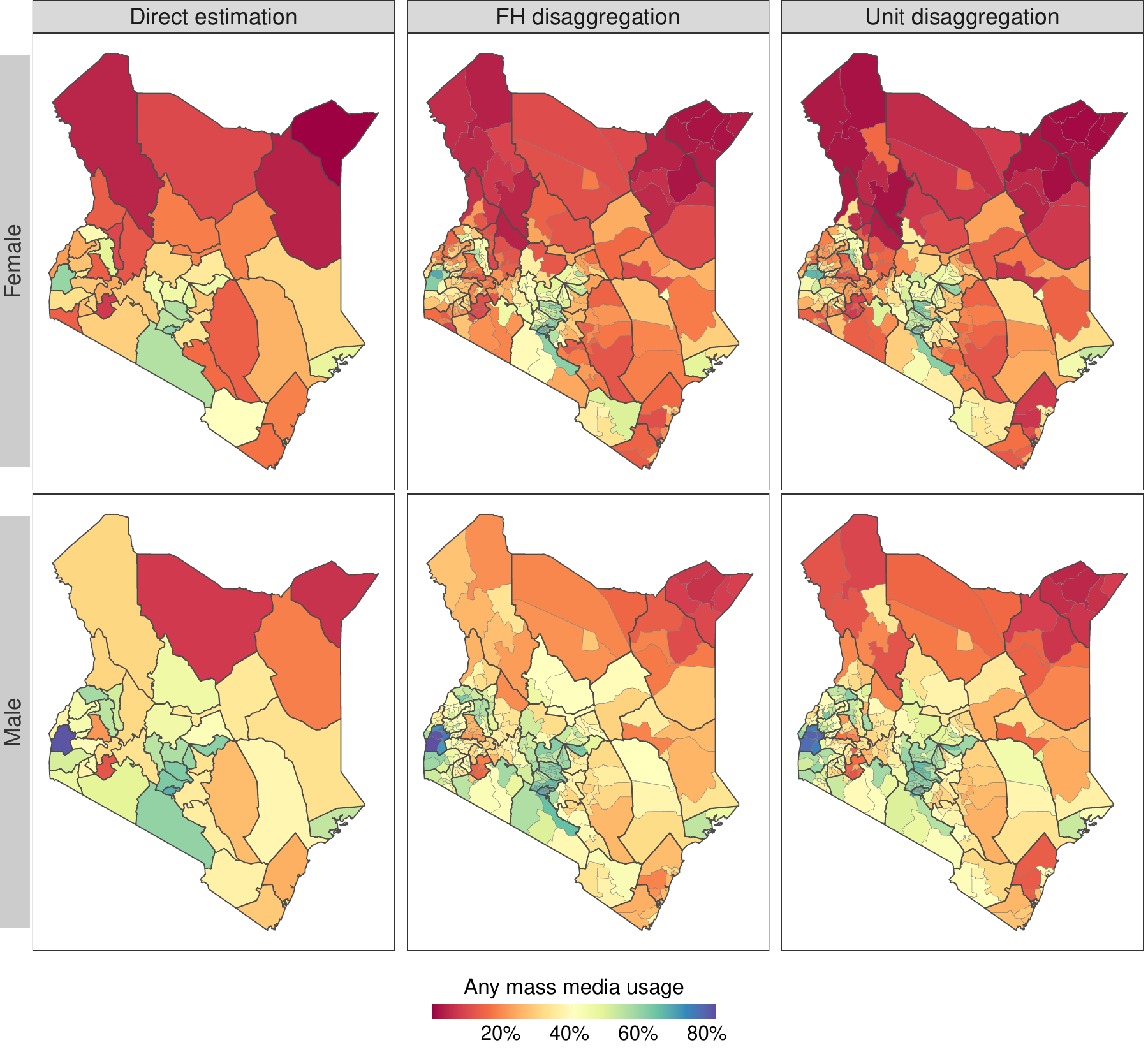}
    \caption{Maps of the female- and male-specific estimates for mass media usage.}
    \label{fig:mass-media-male-female}
\end{figure}

Figure \ref{fig:mass-media-male-female} presents male- and female-specific estimates of the probability of mass media usage across the three modeling approaches. Across Kenya, men report systematically higher usage than women. The spatial pattern shows a trend of Admin-2 regions in Nairobi, Mombasa, and other urban centers having high levels of media use for both sexes, whereas the northern regions show substantially lower rates. Importantly, the disaggregation models recover the within-Admin-2 heterogeneity that the direct Admin-1 estimates cannot capture. Because the indicator is binary and survey samples have higher variability, uncertainty intervals are wider than in previous analyses. As shown in~Figure~\ref{fig:mass-media-male-female-CI} of the supplemental material, credible intervals for disaggregation models reflect this added uncertainty and underscore the need for caution when drawing inferences about fine-scale differences.

\clearpage

\section{Discussion}
\label{sec:discussion}
This work proposed a suite of disaggregation models under the SAE framework to obtain fine-scale areal estimates when the survey data are available only at coarser administrative levels. The methods are built upon two standard SAE approaches---the FH model and unit-level models---and incorporate MRP extensions. This framework is designed to leverage three key sources of information: area-level covariates, spatial smoothing, and demographic compositions to link coarse geo-referenced survey data to latent fields defined at the finer target resolution. Computation is conducted efficiently by using approximate Bayesian inference in \texttt{inlabru}.

The simulation study on GFRs in Kenya mimicked the 2022 DHS and compared disaggregation models geo-indexed at the Admin-1 level with benchmark models geo-indexed at the Admin-2 level. The results provide clear evidence of when disaggregation is effective and when its utility is more limited. When outcome variation is driven by observed covariates or by a mix of observed and spatially correlated unobserved covariates, the models recover within-Admin-1 heterogeneity at the Admin-2 level with strong accuracy. Performance declines sharply, however, when unstructured Admin-2 random effects are introduced or when coefficients vary across strata in ways not accounted for in the models. Uncertainty intervals for disaggregation models are generally wider, and such conservativeness is desirable considering the extrapolative nature of the task. The MRP extension contributes little in terms of predictive power but plays a useful role in the unit-level setting by incorporating survey design. Overall, disaggregation models perform comparably to benchmarks when fine-scale geo-indexing is unavailable. Based on the simulation study, two models emerge as practical choices for real-world application: the FH disaggregation model without MRP and the unit-level disaggregation model with MRP.
%These scenarios highlight a key feature: disaggregation cannot recover variation that is both independent of covariates and devoid of spatial structure. 

The case studies illustrate the practical value of the proposed methods when applied to real survey data. In the DHS 2022 case study, in which Admin-2 identifiers were available for validation, disaggregation models produced estimates largely consistent with benchmark approaches based on Admin-2 indexing. In the 2021 KTUS, disaggregation models are the only viable options because the released data were linked only to Admin-1 areas. In this setting, the models generated meaningful Admin-2 estimates for key indicators such as unpaid domestic and care work and mass media usage. These estimates revealed heterogeneity within counties and highlighted localized gender gaps that would not have been apparent with direct Admin-1 analysis alone. Together, the case studies show that disaggregation can provide valuable subnational insights for localized policy monitoring and SDG reporting, especially where demand for fine-scale estimates is growing. At the same time, the wider uncertainty intervals of the disaggregated results highlight the importance of exercising caution when drawing inferences from such estimates.

We conclude with the limitations of our work. First, disaggregation is fundamentally constrained by the information available. Fine-scale unstructured heterogeneity, represented in the models as area-level IID random effects, cannot be identified from coarsely geo-referenced data. When such unstructured variation dominates the outcome, disaggregation models are expected to perform poorly. Although the simulation study considered a range of plausible scenarios, real-world data-generating processes are likely even more complex. Moreover, for settings in which disaggregation is most relevant, where only coarse geo-indexing is available (e.g., KTUS), direct validation against finer-resolution data is not possible. For this reason, inferences drawn from disaggregation should always be interpreted with caution.

Second, disaggregation models place heavier reliance on covariates than traditional SAE methods applied to sufficiently geo-referenced data. This makes both the choice and the quality of covariates more crucial. In our study, we treated covariates as fixed inputs without explicitly accounting for their uncertainty. Within this disaggregation framework, covariates are used primarily for prediction rather than for inferring causal relationships with the outcomes. As noted in the introduction, the most useful demographic and health covariates typically come from censuses, but census data are infrequent, and their availability is uncertain, particularly in LMICs. 
%Thus, while our applications in Kenya illustrate the feasibility of the approach, the models might not be readily applicable to contexts with weaker covariate availability or less reliable demographic marginals. 
%In practice, covariates derived from modeled data may propagate additional error and risk creating dependencies across data sources. Whenever possible, covariates should be obtained from direct data collection such as census rather than secondary modeling.  Although our framework is predictive rather than causal, avoiding circular dependencies is still preferable. 

Finally, there is room for improvement in the modeling specification. For unit-level models, the most accurate representation of the aggregation process from fine-level latent fields to coarser areas would be a mixture likelihood, where each unit's contribution is aggregated by the probability of belonging to different subareas. However, implementing full Bayesian inference under this specification is computationally demanding. Our approach uses over-dispersed approximations that provide the correct mean representation, which is the primary quantity of interest. Although this approximation may be less accurate in cases of extreme heterogeneity across subareas, there is a significant gain in computational efficiency provided by the \texttt{inlabru} framework. Additionally, our simulation results show no evidence of systematic bias or instability from such approximation.

% On comparing the performance between models that knows admin-2 belonging and models that knows at aggregated level. The advantage of admin-2 model depends on the magnitude of the unstrcutured area iid effect. Because the variability of a region depends on three things, first is measured covariate, second is unmeasured covariate (can be thought of as mostly structured spatial random effect), and the third is completely iid random effect. Knowing the admin-2 belonging is most helpful in identifying the last point, which is completely unobserved at the aggregated level. If this component takes most of the randomness, then admin-2 model will prevail. On the other hand, knowing admin-2 might be harmful in the sense that the model can hard distinguish between true area-level random effects and sample uncertainty such that admin-2 model will encounter this unidentifiability issue such that mistakenly such the randomness from measurement error into the area-level effect. Which might be an issue when the sampling error is so large that we do not want the noise.
% On model with and without spatial random effect. The uncertainty for the model with only covariates can be very low, which omits the spatial random effect. We cannot treat them as measurement error and it is interpreted differently from the indiviudal random effects, which can be integrated out. This mistakenly wrong. Gets back to this point that machine learning prediction can underestimate uncertainty.

\section*{Acknowledgment}
This material is based upon work supported by the US Department of Energy, Office of Science, Advanced Scientific Computing Research program under Award Number DE-SC-ERKJ422.

\bibliographystyle{natbib} 
\bibliography{refs}

\newpage 
\setcounter{section}{0}
\setcounter{figure}{0}
\setcounter{table}{0}
\setcounter{equation}{0}
\setcounter{page}{1}

\makeatletter
\renewcommand \thesection{S\@arabic\c@section}
\renewcommand \thefigure{S\@arabic\c@figure}
\renewcommand \thetable{S\@arabic\c@table}
\renewcommand \theequation{S\@arabic\c@equation}
\renewcommand \thepage{S\@arabic\c@page}
\makeatother

\centerline{{\huge Supplemental Materials for ``Areal Disaggregation:}}
\vspace{3pt}
\centerline{{\huge An SAE Perspective"}}
\vspace{10pt}
%\centerline{\Large Yunhan Wu}

\section{On the Limitations of the Identity Link under \texttt{inlabru} Linearization}
\label{supp:identity-vs-log-link}
It may be tempting to consider an alternative formulation of the unit-level disaggregation model. Because the location parameter $\mu_i$ in the data model is itself the mean of the distribution, it may appear possible to identify it directly with the predictor $\eta_i$, thereby dispensing with the link function that is standard in generalized linear model-based settings. This line of reasoning would suggest that the outer transformations used in the predictor (e.g., $\log(\cdot)$ in $\log(\sum \exp(\cdot))$) could cancel out with the link function in the sampling model. To illustrate why this is incorrect, it is helpful to examine the Poisson case, and the same argument extends to negative binomial and (beta-) binomial likelihoods.

\[
\begin{array}{r@{\quad} | @{\quad} l @{\qquad} l}
\textbf{} 
& \textbf{Problematic} 
& \textbf{Correct} \\ \hline
\text{Sampling model} 
& Y_i \sim \mathrm{Poisson}(n_i\,\mu_i) 
& Y_i \sim \mathrm{Poisson}(n_i\,\mu_i) \\[0.6ex]
\text{Link} 
& \mu_i = \eta_i \quad \text{(identity link)} 
& \log \mu_i = \eta_i \quad \text{(log link)} \\[0.6ex]
\text{Predictor} 
& \eta_i = \displaystyle \sum_{j} \exp (\bm{A}\bm{u}_j)
& \eta_i = \displaystyle \log\!\Big(\sum_{j}  \exp (\bm{A}\bm{u}_j)\Big)
\end{array}
\]

The formulation on the left (above) reflects the apparent but misleading cancellation between the logarithm in the aggregated predictor and the exponential from the log-link. Although algebraically attractive, this setup loses an important functionality of the link: it allows $\eta_i$ to remain unconstrained on $\mathbb{R}$ while ensuring $\mu_i > 0$ for a valid mean. 

Within the \texttt{inlabru} framework, this role is especially critical because predictors such as 
\[
\eta_i \;=\; \log\Big(\sum_{j} \exp (\bm{A}\bm{u}_j)\Big)
\]
are linearized to enable inference. If one were instead to linearize 
\[
\sum_{j} \exp (\bm{A}\bm{u}_j) \;\approx\; \tilde{\bm{A}}\bm{u} + \delta(\bm{u}_0)
\]
and use an identity link, then there would be no guarantee that the approximation remains strictly non-negative across the full support of the latent Gaussian field $\bm{u}$. Although the function value should stay positive at the linearization point, the approximation might result in negative values in the tails of the distribution support, which would be incompatible with the Poisson likelihood (and with the constraints of binomial models).

By contrast, retaining the log-link ensures that even after linearization, the predictor $\eta_i$ remains unconstrained while the transformation $\exp(\eta_i)$ preserves the required positivity of $\mu_i$. Thus, the algebraically simpler formulation cannot be adopted because it breaks the validity of the linearization strategy employed by \texttt{inlabru}.

\newpage

\section{Model Specification for Spatial Model}
\label{sec:BYM2-full-specification}

\textbf{Spatial Effects for Areas:}
The term \( b_j \) represents the spatial random effect for area \( j \), where cluster \( c \) is located. We adopt the BYM2 model \citep{riebler2016an}, which is a reparameterization of the Besag-York-Mollie (BYM) model \citep{besag1991bayesian} and decomposes \( b_j \) into an independent and identically distributed (IID) component \( e_j \) and a spatially structured component \( S_j \):  
\begin{equation*}
b_j = \sigma_b (\sqrt{1-\kappa} e_j + \sqrt{\kappa} S_j),
\end{equation*}
with \(e_j \stackrel{\text{iid}}{\sim} \mathcal{N}(0,1)\) and \(S=(S_1,\ldots,S_n)^\top\) following a scaled intrinsic conditional autoregressive prior  built from the adjacency matrix \(W\) (based on the geographical configuration of areas).
Writing \(j\sim q\) for spatial adjacency and \(m_j\) for the number of neighbors, the conditional for region $j$ is
\[
S_j \,\big|\, \{S_q: q \sim j\} \sim \mathcal{N}\!\left(\frac{1}{m_j}\sum_{q \sim j} S_q,\ \frac{\sigma_s^{2}}{m_j}\right),
\]
with the sum-to-zero constraint \(\sum_j S_j=0\). We use penalized complexity priors \citep{simpson2017penalising} for the hyperparameters $\sigma_b$ (total standard deviation) and $\kappa$ (the proportion of variation that is spatial) such that \(\Pr(\sigma_b>1)=0.01\) and \(\Pr(\kappa>0.5)=0.5\). Our prior specification is predominantly dictated by the default settings in \texttt{R-INLA} and \texttt{inlabru}.

Posterior inference is conducted with the \texttt{inlabru} package \citep{lindgren2024inlabru,bachl2019inlabru}, as discussed in greater detail in Section~\ref{sec:model-implementation-inlabru} of the main manuscript. 

\newpage

\section{Additional Details on Simulation Setup}
\label{sec:supp-simulation-setup}

\subsection{Data Sources and Construction of Inputs}
\label{sec:data-source}

The simulation relies on publicly available geospatial covariates and administrative boundary data to construct a realistic master frame.
First, we obtain analysis shapefiles for Admin-1 and Admin-2 levels from Kenya's Independent Electoral and Boundaries Commission and harmonize names and codes with the census and WorldPop layers to ensure consistent geo-indexing.

Second, we extract sociodemographic marginals from the 2019 Kenya Population and Housing Census at the Admin-2 level, including the proportion urban, the proportion of women with secondary or higher education, and selected additional area-level covariates (e.g., average household size, mobile phone access). Because the census reporting units do not perfectly match the standard Admin-2 boundaries used in our analysis, we harmonize the two by using a crosswalk table that maps census units to analysis units. This mapping is used primarily for transferring fraction or ratio measures. Population counts are taken from WorldPop (described below) and disaggregated using the census proportions. Full details of the crosswalk construction and caveats are provided in the next section.

Third, we use WorldPop 2019's gridded population estimates to obtain counts of females by 5-year age groups on a 1\,km~$\times$~1\,km grid. We aggregate these counts to the Admin-2 level by using the analysis shapefile and produce age-by-area totals for the groups 15--19, ..., 40--44. These totals serve as the base population.

Fourth, we compile additional 1\,km raster covariates to drive spatial heterogeneity in fertility: nighttime lights \citep{roman2018nasa}, travel time to the nearest health facility \citep{weiss2020global}, and vegetation index (NDVI) \citep{didan2015modis}. We then compute population-weighted Admin-2 averages of these covariates.

Fifth, we extract key design parameters from the 2022 Kenya DHS (demographic and health survey): the stratification scheme (Admin-1 $\times$ urban/rural [notably, Nairobi and Mombasa are entirely urban]), number of enumeration areas (EAs) and primary sampling units (PSUs) in the frame by stratum, and the target number of sampled EAs and women per EA. These inputs are used to mirror the two-stage DHS sampling.

\subsection{Census-Admin-2 Crosswalk and Data Harmonization}
\label{sec:cross-walk}

The 2019 Kenya Population and Housing Census reports sociodemographic indicators at the level of 333 subcounties after excluding protected areas and other non-residential units. These reporting units do not perfectly align with the 290 Admin-2 (constituency) boundaries used in our analysis, and no official boundary files for the 333 census subcounties are publicly available. To harmonize these definitions, we constructed a $333 \times 290$ crosswalk matrix linking census subcounties to the Admin-2 units used in our simulation framework.

The crosswalk was developed through a sequential process. We began by matching unit names between the census tables and the Admin-2 shapefile and accounting for known historical name changes, alternative spellings, and differences in formatting. When direct matches were not possible, we manually examined regional records, administrative maps, and auxiliary geographic metadata to identify the most plausible correspondences. For cases in which a census subcounty overlapped multiple Admin-2 units (or vice versa), we permitted many-to-many mappings in the crosswalk. Because such links can distort absolute population counts, we restricted the use of the crosswalk to transferring proportion and ratio measures, such as the fraction of urban or the fraction of women with secondary or higher education.

Absolute counts were instead obtained from UN-adjusted WorldPop 2019 gridded population estimates for 5-year female age groups at a 1\,km resolution. These counts were aggregated to the Admin-2 boundaries used in the analysis and then partitioned by urban/rural and education strata by using the proportions derived from the census. This procedure mitigates the impact of potential census undercounts while preserving the census-derived sociodemographic structure. Although undercounting in census totals has minimal effect on the proportion measures we derived, it could bias counts if used directly, which reinforces our decision to rely on WorldPop data for absolute population totals.\clearpage

\subsection{Generating the Master Frame and the Sample}
\label{sec:master-frame}

\paragraph{Step 1: Individual listing.}
We expand the Admin-2 $\times$ age-group $\times$ urbanicity $\times$ education cells into a synthetic population of individual women from age groups 15--19, 20--24, ..., 40--44. We discuss generation of exposure and outcome in Section~\ref{sec:supp-scenario}.

\paragraph{Step 2: Enumeration areas assignment.}
The DHS report contains data on how many EAs exist within each stratum $h$ (Admin-1 $\times$ urban/rural) in the sampling master frame. We allocate a number of EAs to each Admin-2 $\times$ urban/rural cell proportional to its female population share within each strata. To induce realistic cluster size heterogeneity, we draw EA relative sizes $S$ from a lognormal distribution with $\log S \sim \mathcal{N}(0,\,0.2^2)$. We then partition the individual women in each stratum into EAs according to these relative sizes (proportional allocation) to preserve Admin-2 membership.

\paragraph{Step 3: Two-stage sample selection (mimic DHS).}
We emulate the DHS design by using stratified two-stage sampling. In the first stage, within each stratum $h$, we select $n_h$ clusters (EAs) by probability-proportional-to-size (PPS) sampling, where the measure of size is the number of women in the cluster. Let $N_{hc}$ denote this measure of size for cluster $c$ in stratum $h$ and let $N_{h\cdot}=\sum_c N_{hc}$. The first-stage inclusion probability is
\[
\pi^{(1)}_{hc} = \frac{n_h\,N_{hc}}{N_{h\cdot}} \quad 
\]

In the second stage, from each sampled cluster $c$, we select $m_{hc}$ women by simple random sampling without replacement. We set $m_{hc}=20$ to approximate the sample size in the 2022 Kenya DHS. The conditional probability of selecting woman $k$ in cluster $c$ is then
\[
\pi^{(2)}_{k|hc} = \frac{m_{hc}}{N_{hc}}.
\]
The overall inclusion probability is
\[
\pi_k = \pi^{(1)}_{hc} \times \pi^{(2)}_{k|hc}
      = \frac{n_h\,N_{hc}}{N_{h\cdot}} \times \frac{m_{hc}}{N_{hc}} = \frac{n_h\,m_{hc}}{N_{h\cdot}},
\]
and the base sampling weight is $w_k = 1/\pi_k$. We do not simulate or adjust for nonresponse. 

\subsection{Outcome Generation and Simulation Scenarios}
\label{sec:supp-scenario}

\begin{table}[ht]
\centering
\caption{Sources of spatial heterogeneity under four scenarios and the expected ability of disaggregation models to capture them.}
\renewcommand{\arraystretch}{1.4}
\begin{tabular}{|l|c|c|c|c|c|}
\hline
\textbf{Variation sources} & \textbf{Scenario 1} & \textbf{Scenario 2} & \textbf{Scenario 3} & 
\textbf{Scenario 4} & \textbf{Recoverable} \\ \hline

\makecell[l]{Observed covariates} &
\cmark &
\cmark &
\cmark &
\cmark &
\makecell[l]{\cmark \\ Explicitly captured \\ by fixed effects} \\ \hline

\makecell[l]{Unobserved covariates \\ (spatially correlated \\ Admin-2 effects)} &
\xmark &
\cmark &
\cmark &
\cmark &
\makecell[l]{\opencircle \\ Partially captured \\ by spatial effects} \\ \hline

\makecell[l]{IID Admin-2 \\ random effects} &
\xmark &
\xmark &
\cmark &
\cmark &
\makecell[l]{\xmark \\ Not recoverable, \\ purely unstructured} \\ \hline

\makecell[l]{Varying \\ coefficients} &
\xmark &
\xmark &
\xmark &
\cmark &
\makecell[l]{\xmark \\ Leads to model \\ misspecification} \\ \hline

\end{tabular}
\label{tab:four_simulation_scenarios_supp}
\end{table}

\paragraph{Scenario 1:} In the first scenario, only observed covariates are included. The log-scale mean structure for individual $k$ in cluster $c$ located in Admin-2 area $j$ is specified as
\begin{align*}
\log \mu^{s_1}_{c,k} &=  \alpha_{\mathrm{rural}}\,I(c \in \text{rural})
    + \alpha_{\mathrm{urban}}\,I(c \in \text{urban}) \\
    &\quad + \delta_{\mathrm{age}(k)} + \delta_{\mathrm{educ}(k)} \\
    &\quad + \bm{\beta}\bm{X}_{j[k]} + e_k + e_c.
\end{align*}
The intercepts are fixed at $\alpha_{\mathrm{urban}}=-2.1$ and $\alpha_{\mathrm{rural}}=-1.8$. The terms $\delta_{\mathrm{age}(k)}$ and $\delta_{\mathrm{educ}(k)}$ represent individual-level covariates that are expected to be captured through MRP. The vector $\bm{X}_{j[k]}$ denotes area-level covariates, which include nighttime light, travel time to the nearest health facility, and average household size, all of which are standardized. The covariate effects are specified as
\[
\beta_{\mathrm{ntl}} = -0.08, \quad
\beta_{\mathrm{health}} = 0.08, \quad
\beta_{\mathrm{edu}} = -0.6, \quad
\beta_{\mathrm{hh}} = 0.08.
\]
Education is modeled as a binary indicator of high school attendance, with effect $\delta_{\mathrm{educ}}=-0.6$. The age-group effects are constructed as follows:
\[
\delta_{\mathrm{age}} =
\begin{cases}
-0.3 & \text{if age 15--19},\\
0.6  & \text{if age 20--24},\\
0.5  & \text{if age 25--29},\\
0.3  & \text{if age 30--34},\\
-0.2 & \text{if age 35--39},\\
-0.9 & \text{if age 40--44}.
\end{cases}
\]

These covariate effect values were selected to introduce variability across subgroups and to approximate realistic patterns to some extent. However, they are not intended to serve as direct or accurate estimates of the true underlying effects. The individual-level random effects are drawn from normal distributions with standard deviations $0.05$ for both $e_k$ and $e_c$.

\paragraph{Scenario 2:} In the second scenario, we extend the model by including unobserved covariates in addition to those already specified. The extended structure is
\[
\log(\mu^{s_2}_{c,k}) = \log(\mu^{s_1}_{c,k}) + \bm{\beta}_{\mathrm{unobs}}\bm{X}^{\mathrm{unobs}}_{j[k]},
\]
where $\bm{X}^{\mathrm{unobs}}_{j[k]}$ includes two measures: the vegetation index (NDVI) and female mobile phone usage, both of which are standardized. The effects of these covariates are specified as
\[
\beta_{\mathrm{ndvi}} = 0.08, \quad
\beta_{\mathrm{mobile}} = -0.08.
\]

\paragraph{Scenario 3:} The third scenario builds on the second by introducing an independent random effect at the Admin-2 level:
\[
\log(\mu^{s_3}_{c,k}) = \log(\mu^{s_2}_{c,k}) + e_j, \qquad e_j \sim \mathcal{N}(0, 0.12^2).
\]
This added Admin-2 level IID random effect represents unstructured residual variations that are not expected to be explained by the disaggregation model.

\paragraph{Scenario 4:} In the final scenario, we introduce model misspecification by allowing coefficients to differ between urban and rural parts of the same area. In contrast to the uniform coefficients assumed in the previous scenarios, this specification captures heterogeneity across settlement types. The coefficients for the two strata are as follows:

\noindent\textbf{Urban:} $\alpha=-2.10,\ \beta_{\mathrm{ntl}}=-0.10,\ \beta_{\mathrm{health}}=0.10,\ \beta_{\mathrm{edu}}=-0.50,\ \beta_{\mathrm{hh}}=0.06,\ \beta_{\mathrm{ndvi}}=0.06,\ \beta_{\mathrm{mobile}}=-0.06, \ \delta^{\mathrm{educ}}_k=-0.5$

\noindent\textbf{Rural:} $\alpha=-1.80,\ \beta_{\mathrm{ntl}}=-0.06,\ \beta_{\mathrm{health}}=0.06,\ \beta_{\mathrm{edu}}=-0.70,\ \beta_{\mathrm{hh}}=0.10,\ \beta_{\mathrm{ndvi}}=0.10,\ \beta_{\mathrm{mobile}}=-0.10, \delta^{\mathrm{educ}}_k=-0.7$
\[
\delta^{(\mathrm{Urban})}_{\mathrm{age}} =
\begin{cases}
-0.4 & \text{if age group is 15--19},\\
0.5  & \text{if age group is 20--24},\\
0.5  & \text{if age group is 25--29},\\
0.3  & \text{if age group is 30--34},\\
-0.1 & \text{if age group is 35--39},\\
-0.8 & \text{if age group is 40--44}.
\end{cases}
\qquad
\delta^{(\mathrm{Rural})}_{\mathrm{age}} =
\begin{cases}
-0.2 & \text{if age group is 15--19},\\
0.6  & \text{if age group is 20--24},\\
0.4  & \text{if age group is 25--29},\\
0.3  & \text{if age group is 30--34},\\
-0.2 & \text{if age group is 35--39},\\
-0.9 & \text{if age group is 40--44}.
\end{cases}
\]

\newpage
\section{Additional Simulation Results}
\label{sec:supp-simulation-res}
\begin{table}[!ht]
\caption{Performance metrics across methods and scenarios. Values show the mean for each selected metric across 500 simulation repetitions.}
\centering
\fontsize{12pt}{14pt}\selectfont
\begin{tabular}{lrrrrr}
\toprule
\textbf{Method} & \textbf{Regression $R^2$}& \textbf{Pearson r} & \textbf{Interval Score} & \textbf{Bias} & \textbf{Abs Rel Bias} \\
\midrule
\multicolumn{6}{l}{\textbf{Scenario 1}} \\
\cmidrule(lr){1-6}
FH disagg         & 0.927 & 0.978 & 0.014 &  0.000 &  3.6\% \\
FH-MRP disagg     & 0.929 & 0.982 & 0.014 &  0.000 &  3.7\% \\
Unit disagg       & 0.949 & 0.983 & 0.011 &  0.000 &  3.1\% \\
Unit-MRP disagg   & 0.983 & 0.997 & 0.009 &  0.000 &  1.7\% \\
FH (Adm2)         & 0.642 & 0.832 & 0.020 &  0.000 &  7.6\% \\
Direct (Adm2)     & 0.062 & 0.690 & 0.029 &  0.000 & 13.3\% \\
\hline
\addlinespace[4pt]
\multicolumn{6}{l}{\textbf{Scenario 2}} \\
\cmidrule(lr){1-6}
FH disagg         & 0.857 & 0.962 & 0.024 &  0.000 &  5.5\% \\
FH-MRP disagg     & 0.821 & 0.957 & 0.025 & -0.003 &  6.3\% \\
Unit disagg       & 0.911 & 0.979 & 0.021 &  0.000 &  4.2\% \\
Unit-MRP disagg   & 0.902 & 0.976 & 0.020 & -0.001 &  4.3\% \\
FH (Adm2)         & 0.691 & 0.859 & 0.021 &  0.001 &  8.1\% \\
Direct (Adm2)     & 0.291 & 0.738 & 0.029 &  0.001 & 13.4\% \\
\hline
\addlinespace[4pt]
\multicolumn{6}{l}{\textbf{Scenario 3}} \\
\cmidrule(lr){1-6}
FH disagg         & 0.622 & 0.773 & 0.034 &  0.000 & 10.8\% \\
FH-MRP disagg     & 0.598 & 0.770 & 0.036 & -0.002 & 11.2\% \\
Unit disagg       & 0.670 & 0.787 & 0.031 &  0.000 & 10.0\% \\
Unit-MRP disagg   & 0.668 & 0.786 & 0.032 &  0.000 & 10.0\% \\
FH (Adm2)         & 0.685 & 0.839 & 0.023 &  0.001 &  9.9\% \\
Direct (Adm2)     & 0.469 & 0.792 & 0.029 &  0.001 & 13.4\% \\
\hline
\addlinespace[4pt]
\multicolumn{6}{l}{\textbf{Scenario 4}} \\
\cmidrule(lr){1-6}
FH disagg         & 0.593 & 0.756 & 0.035 &  0.000 & 10.9\% \\
FH-MRP disagg     & 0.570 & 0.753 & 0.037 & -0.002 & 11.4\% \\
Unit disagg       & 0.641 & 0.771 & 0.032 &  0.000 & 10.3\% \\
Unit-MRP disagg   & 0.628 & 0.764 & 0.034 &  0.000 & 10.4\% \\
FH (Adm2)         & 0.670 & 0.834 & 0.023 &  0.001 &  9.9\% \\
Direct (Adm2)     & 0.449 & 0.789 & 0.028 &  0.001 & 13.4\% \\
\bottomrule
\end{tabular}
\label{tab:sim-res-table}
\end{table}

\newpage 

\begin{figure} [!ht]
    \centering
    \includegraphics[clip, trim=0cm 0cm 0cm 0cm, width=1\linewidth]{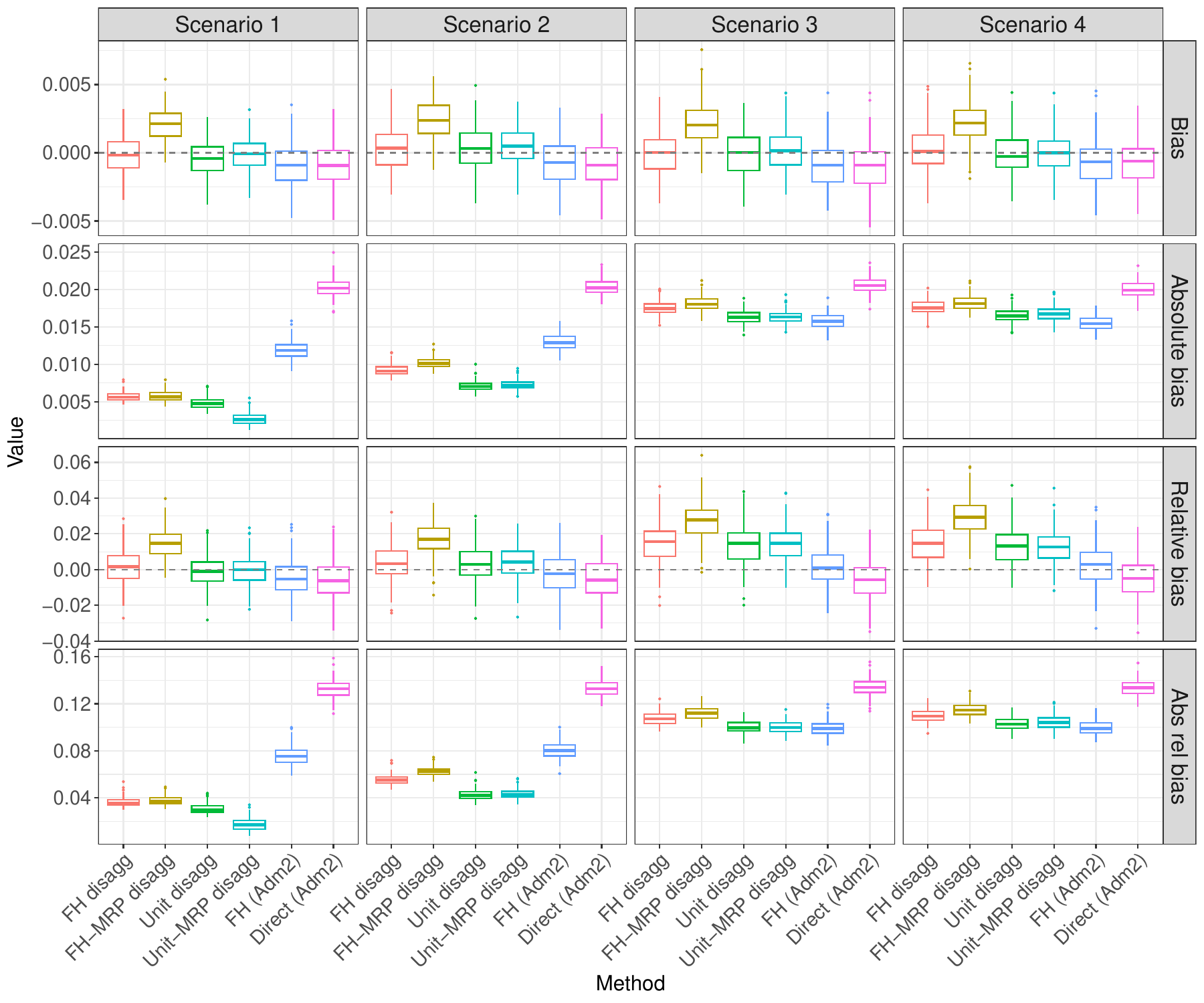}
    \caption{Comparison of disaggregation and benchmark models across four simulation scenarios evaluated with bias metrics. }
    \label{fig:sim-res-box-bias}
\end{figure}

\newpage
\section{Additional Visualizations}

\subsection{Further Comparison of GFR Estimates (Case Study 1)}

\label{sec:supp-case-study-1}

\begin{figure} [!ht]
    \centering
    \includegraphics[clip, trim=0cm 0cm 0cm 0cm, width=1\linewidth]{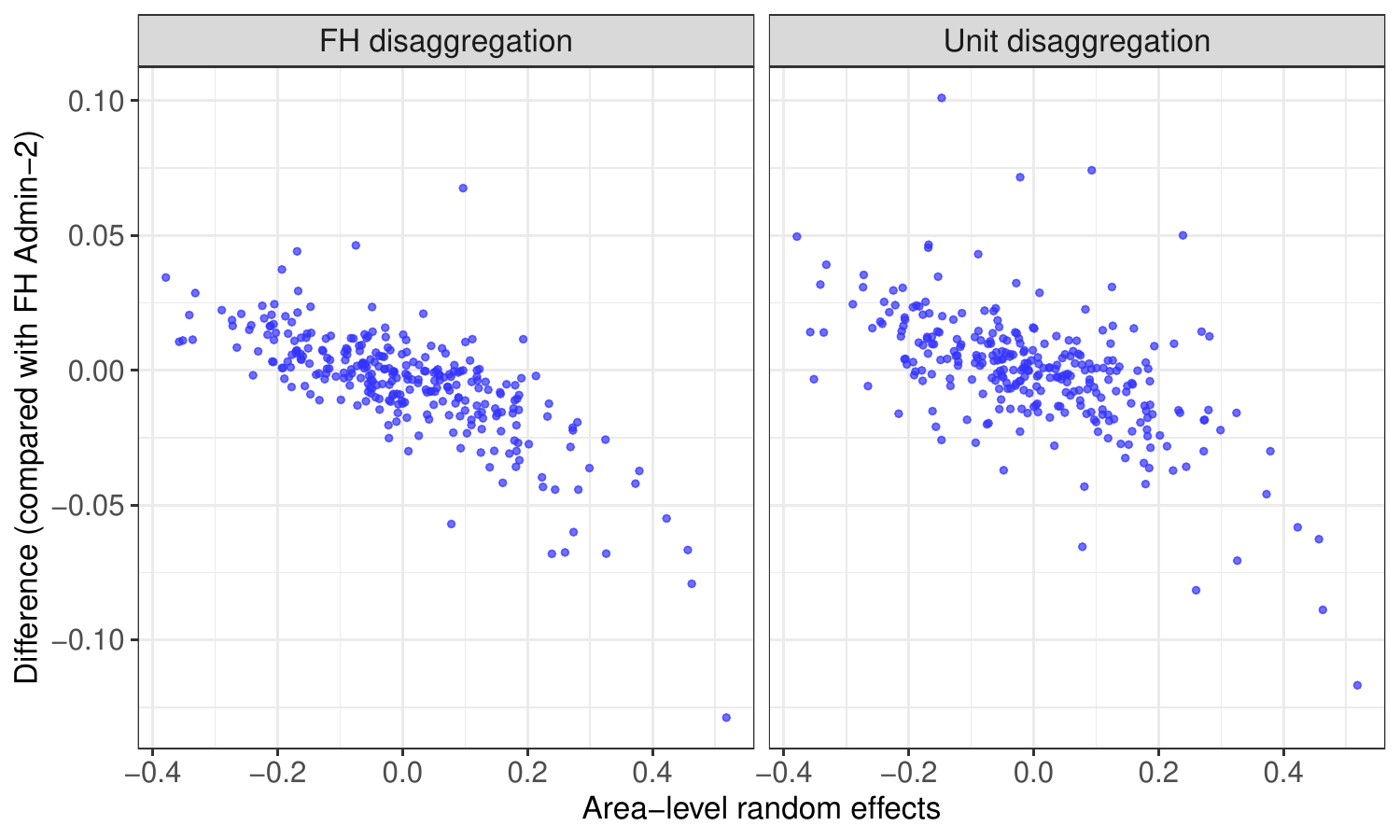}
    \caption{Scatter plots of the relationship between area-level random effects and the difference in GFR estimates from disaggregation models compared with the FH Admin-2 benchmark.}
    \label{fig:scatter-GFR}
\end{figure}

Figure \ref{fig:scatter-GFR} shows the difference between Admin-2 estimates from each disaggregation model and those from the Admin-2 Fay--Herriot (FH) model against the area-level random effects estimated by the Admin-2 FH model. The strong association indicates that the deviation of disaggregation models arises in regions with a strong area-level effect. This is consistent with our simulation findings: by design, disaggregation models have a limited ability to identify area-level random effects, especially the unstructured component.

\newpage

\begin{figure} [!ht]
    \centering
    \includegraphics[clip, trim=0cm 0cm 0cm 0cm, width=1\linewidth]{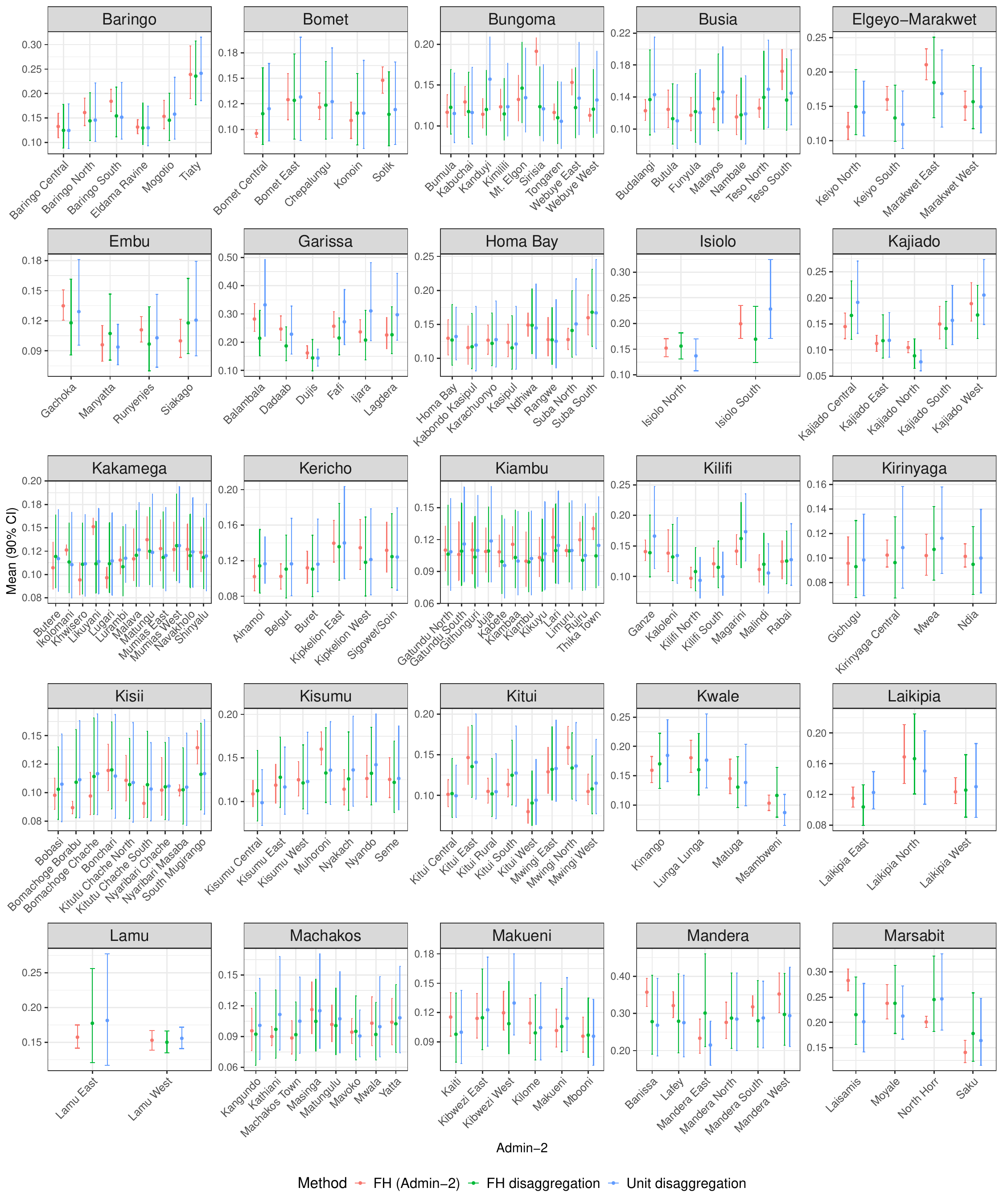}
    \caption{Comparison of Admin-2 point estimates and 90\% credible intervals for GFR across the benchmark FH Admin-2 model and the FH and unit-level disaggregation models displayed within Admin-1 groupings.}
    \label{fig:interval-GFR-1}
\end{figure}

% \begin{figure*} [!ht]
%     \centering
%     \includegraphics[clip, trim=0cm 23.2cm 0cm 0cm, width=1\linewidth]{figs/appendix/compare_methods_interval_plot.pdf}
%     \caption{benchmark.}
%     \label{fig:interval-GFR-1}
% \end{figure*}

% \newpage 

% \begin{figure*} [!ht]
%     \centering
%     \includegraphics[clip, trim=0cm 0cm 0cm 53cm, width=1\linewidth]{figs/appendix/compare_methods_interval_plot.pdf}
%     \caption{benchmark.}
%     \label{fig:interval-GFR-1}
% \end{figure*}

\newpage

\begin{figure} [!ht]
    \centering
    \includegraphics[clip, trim=0cm 0cm 0cm 0cm, width=1\linewidth]{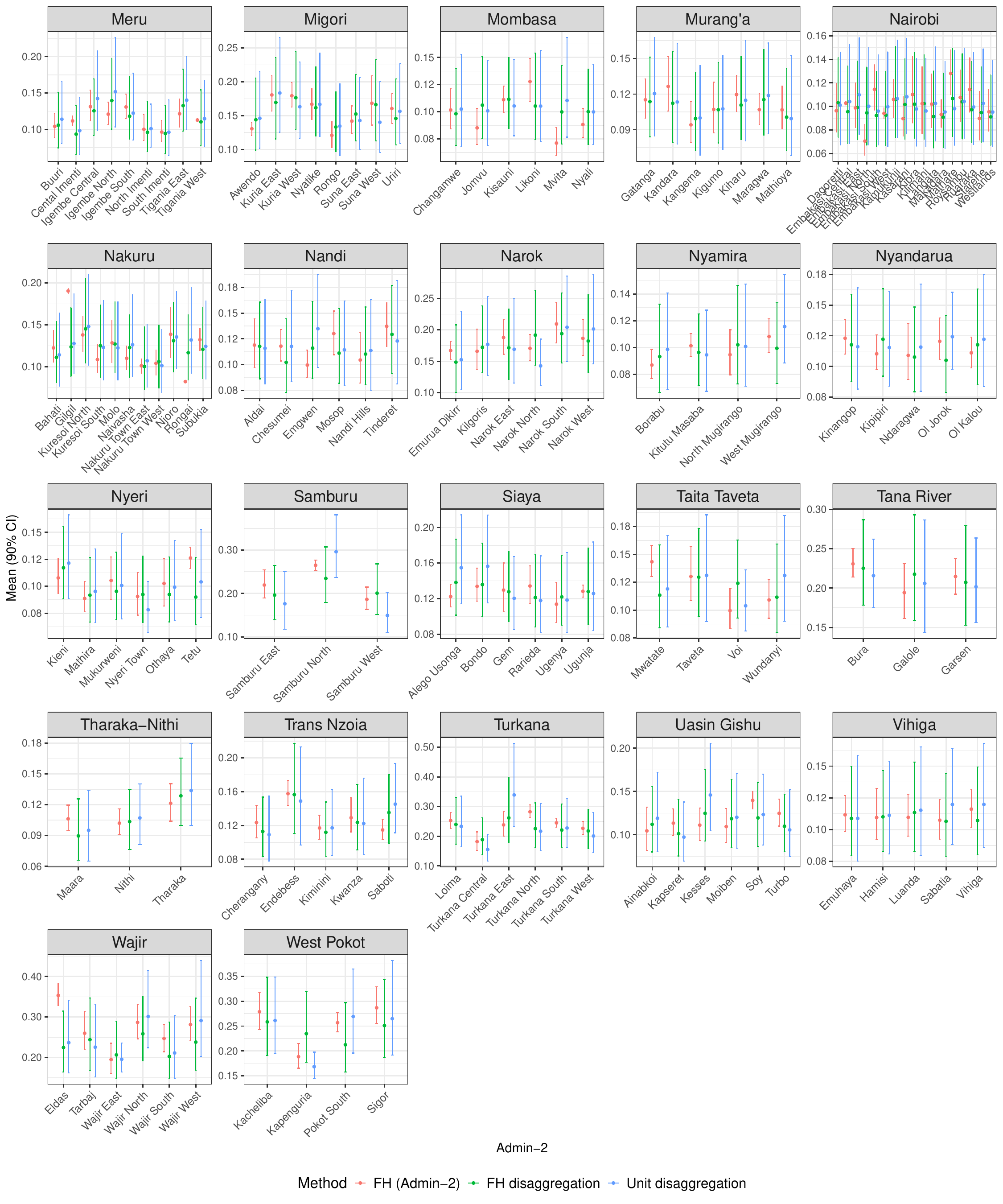}
    \caption{Interval plot continued.}
    \label{fig:interval-GFR-2}
\end{figure}

\newpage

\subsection{Time Spent on Unpaid Care and Domestic Work (Men)}
\begin{figure} [!ht]
    \centering
    \includegraphics[clip, trim=0cm 0cm 0cm 0cm, width=1\linewidth]{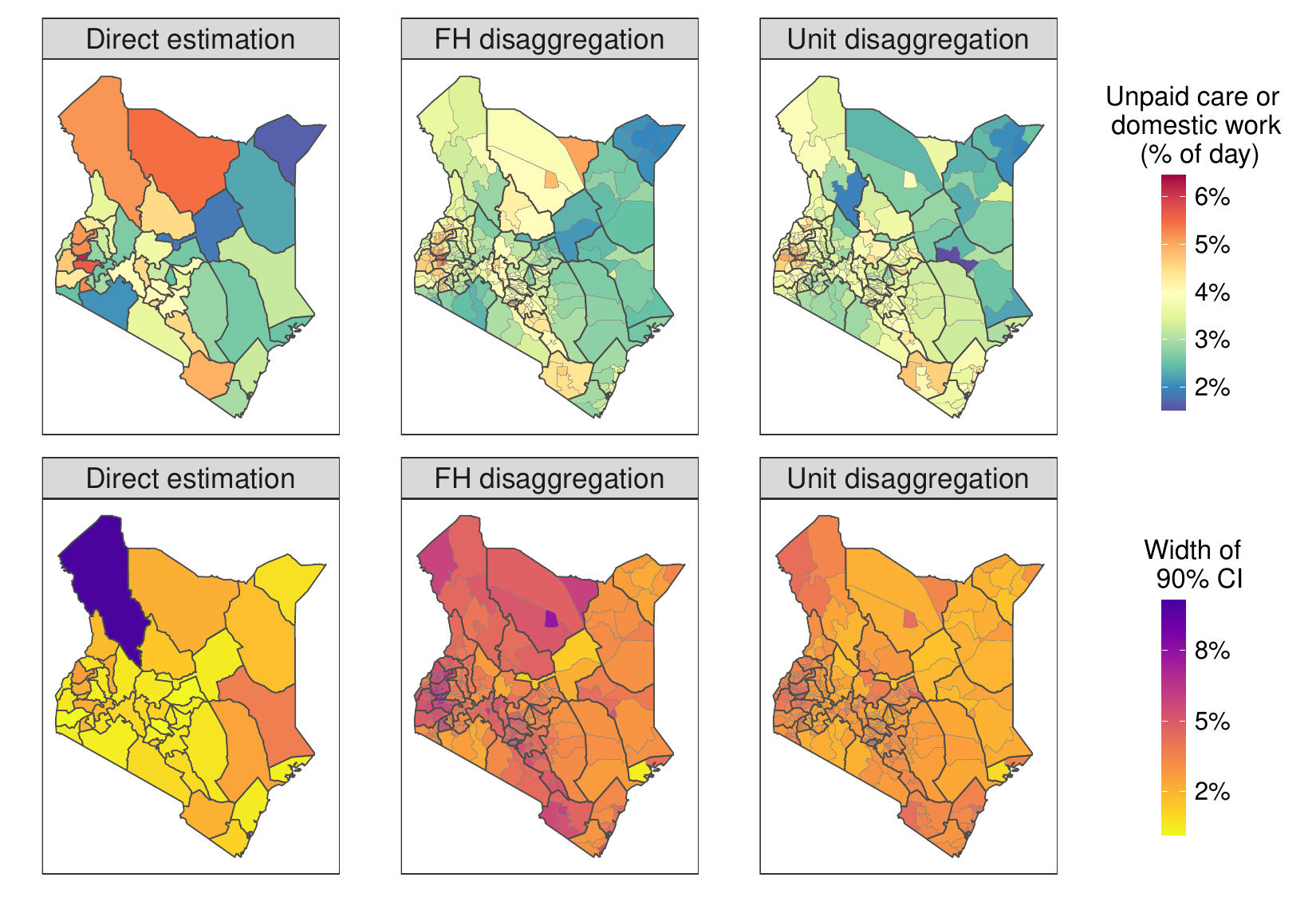}
    \caption{Estimated proportion of the day that men spend on unpaid care and domestic work based on direct estimation at Admin-1, FH disaggregation, and unit-level disaggregation models. The maps display both point estimates and the widths of the associated 90\% uncertainty intervals.}
    \label{fig:male-unpaid-work}
\end{figure}

\newpage 

\subsection{Uncertainty for Male and Female Mass Media Usage Estimates}

\begin{figure} [!ht]
    \centering
    \includegraphics[clip, trim=0cm 0cm 0cm 0cm, width=1\linewidth]{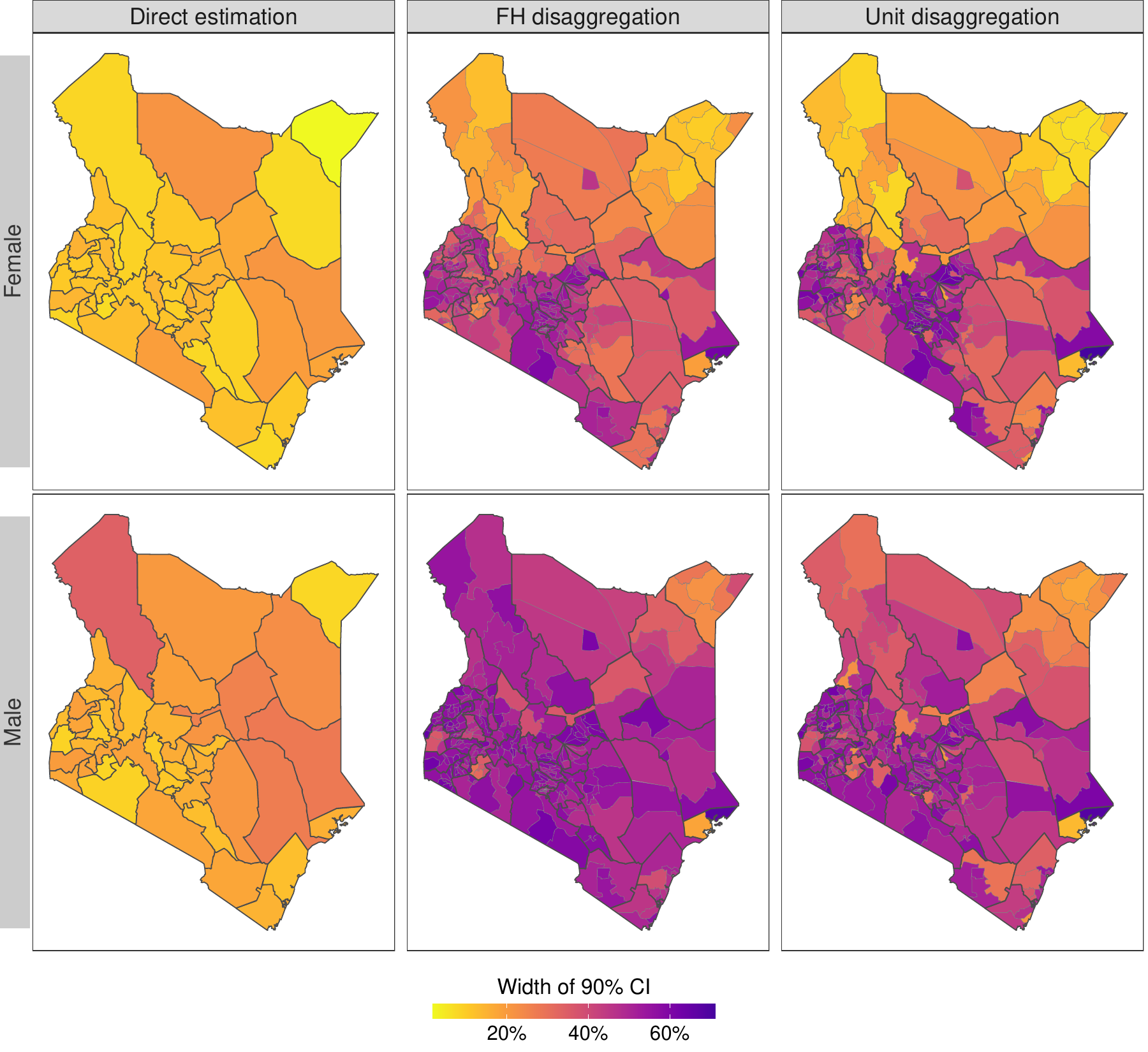}
    \caption{Maps of 90\% CI widths for male and female mass media usage estimates.}
    \label{fig:mass-media-male-female-CI}
\end{figure}

\end{document}